\documentclass[aps,prb,twocolumn,10pt,superscriptaddress,showpacs]{revtex4-1}

\def\bra#1{\mathinner{\langle{#1}|}}
\def\ket#1{\mathinner{|{#1}\rangle}}
\def\braket#1{\mathinner{\langle{#1}\rangle}}

\let\protect\relax
{\catcode`\|=\active
  \xdef\Braket{\protect\expandafter\noexpand\csname Braket \endcsname}
  \expandafter\gdef\csname Braket \endcsname#1{\begingroup
     \ifx\SavedDoubleVert\relax
       \let\SavedDoubleVert\|\let\|\BraDoubleVert
     \fi
     \mathcode`\|32768\let|\BraVert
     \left\langle{#1}\right\rangle\endgroup}
}
\def\BraVert{\@ifnextchar|{\|\@gobble}
     {\egroup\,\mid@vertical\,\bgroup}}
\def\BraDoubleVert{\egroup\,\mid@dblvertical\,\bgroup}
\let\SavedDoubleVert\relax

{\catcode`\|=\active
  \xdef\set{\protect\expandafter\noexpand\csname set \endcsname}
  \expandafter\gdef\csname set \endcsname#1{\mathinner
        {\lbrace\,{\mathcode`\|32768\let|\midvert #1}\,\rbrace}}
  \xdef\Set{\protect\expandafter\noexpand\csname Set \endcsname}
  \expandafter\gdef\csname Set \endcsname#1{\left\{%
     \ifx\SavedDoubleVert\relax \let\SavedDoubleVert\|\fi
     \:{\let\|\SetDoubleVert
     \mathcode`\|32768\let|\SetVert
     #1}\:\right\}}
}
\def\midvert{\egroup\mid\bgroup}
\def\SetVert{\@ifnextchar|{\|\@gobble}
    {\egroup\;\mid@vertical\;\bgroup}}
\def\SetDoubleVert{\egroup\;\mid@dblvertical\;\bgroup}

\begingroup
 \edef\@tempa{\meaning\middle}
 \edef\@tempb{\string\middle}
\expandafter \endgroup \ifx\@tempa\@tempb
 \def\mid@vertical{\middle|}
 \def\mid@dblvertical{\middle\SavedDoubleVert}
\else
 \def\mid@vertical{\mskip1mu\vrule\mskip1mu}
 \def\mid@dblvertical{\mskip1mu\vrule\mskip2.5mu\vrule\mskip1mu}
\fi

\usepackage{graphicx}
\usepackage[T1]{fontenc}
\usepackage[]{amsfonts}
\usepackage[]{amsmath}
\usepackage[]{rotating}
\usepackage[]{color}
\usepackage[]{float}
\usepackage[]{fancyhdr}
\usepackage[]{booktabs}
\usepackage[]{epsfig}
\usepackage{appendix}
\usepackage{amssymb}
\usepackage{psfrag}
\usepackage{setspace}

\begin{document}

\title{Superadiabatic theory for Cooper pair pumping under decoherence}

\author{Juha Salmilehto}
\affiliation{Department of Applied Physics/COMP, Aalto University, P.O. Box 14100, FI-00076 AALTO, Finland}
\author{Mikko M\"ott\"onen}
\affiliation{Department of Applied Physics/COMP, Aalto University, P.O. Box 14100, FI-00076 AALTO, Finland}
\affiliation{Low Temperature Laboratory, Aalto University, P.O. Box 13500, FI-00076 AALTO, Finland}


\begin{abstract}

We introduce a method where successive coordinate transformations are applied to decrease the error in the adiabatic master equation resulting from truncation in the local adiabatic parameter. Our method reduces the non-physical behaviour stemming from the lack of complete positivity. The strong environment-induced relaxation at high Cooper pair pumping frequencies leads to adiabatic ground-state pumping only in the lowest-order approximation. We illustrate the robustness of the frequency where the adiabaticity breaks down using the high-order theory and show the emergence of an optimal environmental coupling strength, for which ideal pumping is preserved for the highest frequency. Finally, we study the effect of quantum interference on the pumped current and give an estimate for the relaxation rate of an experimentally measured system.

\end{abstract}

\maketitle

\section{\label{sec:intro}Introduction}

Detection and manipulation of geometric phases \cite{prsla392/45, prl51/2167, prl52/2111, prl58/1593, prl60/2339} in superconducting quantum devices has been an area of active research in recent years with one of the ultimate goals being the ability to realize holonomic quantum gates \cite{pla264/94}. Even though alternative methods to experimentally generate and detect the geometric phases in such devices have been proposed and realized \cite{nature407/355, science318/1889}, the link they have to Cooper pair pumping has attracted major interest \cite{prb60/R9931, prb63/132508, prb68/020502(R), prl91/177003, prl95/256801, apl90/082102, prl98/127001, prl100/027002, prl100/117001, prb77/144522} as it provides means of detecting the phases by a measurement of the dynamically and geometrically transferred charges. A measurement scheme based on one superconducting island with two tunable Josephson junctions, the Cooper pair sluice in a superconducting loop, has been introduced \cite{prb73/214523} and experimentally realized \cite{prl100/177201}. Recently, a system based on a similar device structure has been proposed to execute fully geometric quantum computing using non-Abelian phases \cite{prb81/174506, pra82/052304}.

Even though the principles of operation for Cooper pair pumps are well-known, accounting for system--environment interactions has been a work in progress. Cooper pair pumping being essentially a coherent process, a proper description of the operation must include the effect of the external environment. The usual methods for describing the dynamics of open steered quantum systems \cite{tToOQS, API, prl94/070407} have been shown to violate the conservation of physical observables such as the electric charge and a variety of means have been employed in an attempt to properly describe the dynamics \cite{pra73/052304, pra73/022327, prl90/160402, prb77/115322, pra82/052107, pra83/012112, pra81/022117, pra83/032122}. Recently, it was discovered that for a consistent description using the master equation for the reduced density matrix of the system, all the non-secular terms must be included to enable relaxation to a proper basis and to ensure conservation of the pumped charge in Cooper pair pumping \cite{prl105/030401, prb82/134517}. The same master equation has also been derived using superadiabatic bases \cite{pra82/062112}.

In this paper, we employ the recent methodology of superadiabatic bases \cite{pra82/062112} in deriving the master equation for steered systems and apply it after multiple coordinate transformations of the time-local basis. This results in a master equation where the truncation error related to the local adiabatic parameter is potentially decreased as a function of the number of coordinate transformations. We apply our method to the problem of Cooper pair pumping. We show that for the zero-temperature environment and fast pumping, the ground-state adiabatic evolution revives in the relaxation dominated region only using the adiabatic basis, that is, in the lowest order of our description. Furthermore, we show that the overestimation of the pumped charge caused by the non-positivity of the density matrix, is alleviated dramatically by the usage of high-order bases. We simulate the breakdown of adiabaticity with increasing pumping frequency and show the emergence of an optimal coupling strength preserving ideal pumping up to the highest frequency. We present a condition for the highest transition probability caused by constructive interference between driving-induced excitations generated at different times and show that it corresponds to the downward resonance peaks in the pumped current. Finally, we obtain an estimate for the relaxation rate of the device employed in the experiments of Ref.~\onlinecite{prl100/177201} to pump Cooper pairs.

The structure of this paper is as follows. In the next section, we introduce the model describing a driven quantum system and demonstrate our method of defining successive effective Hamiltonians by coordinate transformations. In Sec.~\ref{sec:master}, we write a master equation for the matrix elements of the reduced density matrix of the system taken in an $n$-times transformed time-dependent basis. In Sec.~\ref{sec:sluice}, we use the master equation to model Cooper pair pumping. Furthermore, we simulate previous experiments on the pumped current and derive an estimate for the relaxation rate of a measured superconducting system. We conclude the paper in Sec.~\ref{sec:conclusions}.

\section{\label{sec:model}Model}

We study a quantum system with a Hamiltonian $\hat{H}_S$ which depends on a set of real control parameters $\{q_k\}$ that vary in time. The system is assumed to be interacting with its environment such that the total Hamiltonian is
\begin{equation}
\hat{H}(t) = \hat{H}_S(t) + \hat{V}(t) + \hat{H}_E,
\label{eq:H}
\end{equation}
where $\hat{V}(t)$ is the coupling between the system and its environment and $\hat{H}_E$ is the Hamiltonian of the environment. We assume that the coupling is of the generic form $\hat{V} = \hat{A} \otimes \hat{X}(t)$, where $\hat{A}$ is the system part of the coupling operator and $\hat{X}(t)$ acts in the Hilbert space of the environment. Let $\ket{m;\vec{q}(t)}$ be the instantaneous eigenstate of $\hat{H}_S(t)$ and $E_m(t)$ the corresponding eigenenergy defined by $\hat{H}_S[\vec{q}(t)] \ket{m;\vec{q}(t)} = E_m[\vec{q}(t)]\ket{m;\vec{q}(t)}$. In the context of adiabatic evolution, $\{\ket{m;\vec{q}(t)}\}$ is referred to as the \textit{adiabatic basis}. We assume that the adiabatic states are normalized and non-degenerate.

Let the Hamiltonian $\hat{H}_S(t)$ be diagonalized in a fixed time-independent basis $\{\ket{m_f}\}$ using the eigendecomposition as $\hat{\tilde{H}}^{(1)}_S(t)=\hat{D}_1^{\dagger}(t)\hat{H}_S(t)\hat{D}_1(t)$, implying that $\braket{n_f|\hat{\tilde{H}}^{(1)}_S(t)|m_f} = E_m(t) \delta_{nm}$. We define a similar transformation for the total density operator $\hat{\rho}(t)$ in the Schr\"odinger picture as $\hat{\tilde{\rho}}^{(1)}(t)=\hat{D}_1^{\dagger}(t)\hat{\rho}(t)\hat{D}_1(t)$. It follows from the Schr\"odinger equation that the evolution of $\hat{\tilde{\rho}}^{(1)}(t)$ is governed by the effective Hamiltonian for the adiabatic basis
\begin{equation}
\hat{\tilde{H}}^{(1)}(t) = \hat{\tilde{H}}^{(1)}_S(t) + \hbar \hat{w}_1(t) + \hat{\tilde{V}}^{(1)}(t) + \hat{H}_E,
\label{eq:Heff}
\end{equation}
where $\hat{\tilde{V}}^{(1)}(t)=\hat{D}_1^{\dagger}(t)\hat{V}(t)\hat{D}_1(t)=\hat{D}_1^{\dagger}(t)\hat{A}\hat{D}_1(t) \otimes \hat{X}(t)$ and $\hat{w}_1(t) = -i\hat{D}_1^{\dagger}(t)\dot{\hat{D}}_1(t)$. The eigenbasis of $\hat{\tilde{H}}^{(1)}_S(t) + \hbar \hat{w}_1(t)$ is usually referred to as the \textit{superadiabatic basis}.

We can further define a unitary transformation $\hat{D}_2(t)$ making $\hat{\tilde{H}}^{(1)}_S(t) + \hbar \hat{w}_1(t)$ diagonal in the fixed basis \cite{pra82/062112}. Thus the evolution of the density matrix $\hat{\tilde{\rho}}^{(2)} = \hat{D}_2^{\dagger}\hat{\tilde{\rho}}^{(1)}\hat{D}_2 = \hat{D}_2^{\dagger}\hat{D}_1^{\dagger}\hat{\rho}\hat{D}_1\hat{D}_2$ is governed by the effective Hamiltonian for the first superadiabatic basis
\begin{equation}
\hat{\tilde{H}}^{(2)}(t) = \hat{\tilde{H}}^{(2)}_S(t) + \hbar \hat{w}_2(t) + \hat{\tilde{V}}^{(2)}(t) + \hat{H}_E,
\label{eq:Heff2}
\end{equation}
where $\hat{\tilde{H}}^{(2)}_S(t)=\hat{D}_2^{\dagger}(t)[\hat{\tilde{H}}^{(1)}_S(t)+\hbar \hat{w}_1(t)]\hat{D}_2(t)$, $\hat{\tilde{V}}^{(2)}(t)=\hat{D}_2^{\dagger}(t)\hat{\tilde{V}}^{(1)}(t)\hat{D}_2(t)$, and $\hat{w}_2=-i\hat{D}_2^{\dagger}(t)\dot{\hat{D}}_2(t)$. This method of successive coordinate transformations can be continued to yield for the $(n-1)$th superadiabatic basis an effective Hamiltonian of 
\begin{equation}
\hat{\tilde{H}}^{(n)} = \hat{\tilde{H}}_S^{(n)} + \hbar \hat{w}_{n} + \hat{\tilde{V}}^{(n)} + \hat{H}_E,
\label{eq:nH}
\end{equation}
where $\hat{\tilde{H}}_S^{(n)} = \hat{D}_{n}^{\dagger}[\hat{\tilde{H}}_S^{(n-1)} + \hbar \hat{w}_{n-1}]\hat{D}_{n}$, $\hat{\tilde{V}}^{(n)} = (\prod_{i=2}^{n}\hat{D}_i)^{\dagger}\hat{D}_1^{\dagger}\hat{V}\hat{D}_1(\prod_{i=2}^{n}\hat{D}_i)$ and $\hat{w}_{n} = -i\hat{D}_{n}^{\dagger}\dot{\hat{D}}_{n}$, where we omitted explicitly marking the temporal dependence of the operators for clarity. The operator product is defined as $\prod_{i=2}^{n}\hat{D}_i = \hat{D}_2\hat{D}_3 \cdots \hat{D}_{n-1}\hat{D}_{n}$. If we define $\hat{D}_S^{(n)}=\prod_{i=2}^{n}\hat{D}_{i}$ for $n \geq 2$ and $\hat{D}_S^{(n)}=\hat{I}$ for $n=1$, the density operators governed by the Hamiltonians in Eqs.~(\ref{eq:Heff}), (\ref{eq:Heff2}) and (\ref{eq:nH}) obtain a more universal form $\hat{\tilde{\rho}}^{(n)} = (\hat{D}_S^{(n)})^{\dagger}\hat{D}_1^{\dagger}\hat{\rho}\hat{D}_1\hat{D}_S^{(n)}$. Defining successive diagonalizations in this manner proves useful in Sec.~\ref{sec:master} as the recently derived master equation \cite{prl105/030401, prb82/134517, pra82/062112} can be applied to solve the system dynamics using these high-order bases.

The iterative method described here is an adaptation of Berry's concept \cite{prsla414/31} he later referred to as \textit{adiabatic renormalization} \cite{GPIPberry}. It is based on the idea that each transformation rotates the basis we use to describe the system dynamics ever closer to the exact evolving closed system state, that is, the time-dependence of the transformed system Hamiltonian is suppressed after each rotation. After $n$ transformations, we define the time-dependent basis as $\{\hat{D}_1\hat{D}_S^{(n)}\ket{m_f}\}$. This approach generally works only in the restricted sense, that is, after a number of iterations, the following rotations will not allow us to describe the dynamics of the system more accurately \cite{prsla414/31}.

Finally, we introduce the local adiabatic parameter as $\alpha_1(t) = \hbar ||\hat{w}_1(t)||/\Delta(t)$, where we compare the Hilbert-Schmidt norm of the operator arising from the adiabatic evolution $||\hat{w}_1(t)|| = \sqrt{\textrm{Tr}_S\{\hat{w}_1(t)^{\dagger}\hat{w}_1(t)\}}$ to an instantaneous minimum energy gap in the spectrum $\Delta(t)$. Here $\textrm{Tr}_S$ denotes the trace over the system degrees of freedom and in the following we will use $\mathrm{Tr}_E$ to denote the trace over the environment degrees of freedom. The parameter $\alpha_1(t)$ should give a good estimate for the degree of adiabaticity of the evolution \cite{prb82/134517, pra82/062112}. In cyclic evolution with the period $T$, the parameter scales as $1/T$ and, thus, in adiabatic evolution we should have $\alpha_1(t) \ll 1$.

\section{\label{sec:master}Master equation}

We consider an adiabatically steered two-level quantum system weakly coupled to its environment. We denote the ground and excited states of $\hat{H}_S$ in the Schr\"odinger picture as $\ket{g}$ and $\ket{e}$, respectively, with corresponding eigenenergies $E_g$ and $E_e$. Using the interaction picture approach, a master equation was derived to describe the dynamics of such a system in Refs.~[\onlinecite{prl105/030401, prb82/134517, pra82/062112}] up to the linear order in $\alpha_1(t)$ and the quadratic order in the system--environment coupling. The method of derivation employed in Ref.~\onlinecite{pra82/062112} is our starting point for developing a numerical scheme for obtaining a more accurate description of the dissipative system dynamics. In Ref.~\onlinecite{pra82/062112}, a master equation for nonsteered systems was used in conjunction with the effective Hamiltonian in Eq.~(\ref{eq:Heff2}) to derive the leading-order master equation under steering. However, a similar derivation can be carried out using $\hat{\tilde{H}}^{(n)}$ for any $n$ to obtain a master equation for the matrix elements of $\hat{\tilde{\rho}}^{(n)}$. Notice that even though the method of defining successive coordinate transformations can be applied to a system with arbitrary number of energy levels, we constrain ourselves to the two-level case. This is practical since our main goal is to explore the implications of applying our scheme compared to previous results \cite{prl105/030401, prb82/134517}.

We define the reduced density operator of the system as $\hat{\tilde{\rho}}_S^{(n)} = \textrm{Tr}_E\{\hat{\tilde{\rho}}^{(n)}\}$ so that its diagonal element becomes $\rho^{(n)}_{gg} = \braket{0|\hat{\tilde{\rho}}_S^{(n)}|0}$ and the off-diagonal element $\rho^{(n)}_{ge} = \braket{0|\hat{\tilde{\rho}}_S^{(n)}|1}$, where $\{\ket{m_f=0},\ket{m_f=1}\}$ is the relevant fixed basis. These are simply the matrix elements of the usual density operator of the system in the Schr\"odinger picture taken in a time-dependent basis $\{\ket{g^{(n)}}, \ket{e^{(n)}}\}$, where $\ket{g^{(n)}} = \hat{D}_1\hat{D}_S^{(n)}\ket{0}$ and $\ket{e^{(n)}} = \hat{D}_1\hat{D}_S^{(n)}\ket{1}$. This is the rotated basis obtained through the iterative procedure we described in Sec.~\ref{sec:model}. We emphasize that the basis states are not obtained using a perturbative expansion in the local adiabatic parameter and, thus, each iteration generally alters them by terms of all orders of $\alpha_1 (t)$ \cite{prsla414/31, pra78/052508}. However, if the method described in Ref.~\onlinecite{pra82/062112} is applied in the $n$-times transformed basis, the error in the resulting master equation is defined by a perturbative expansion in $\hat{w}_n$. The norm of $\hat{w}_n$ decreases, in the restricted sense, with increasing $n$ as the time-dependence of the transformed system Hamiltonian is suppressed. In regard to depicting the actual dynamics of the system, the only issue relevant to the selection of the basis is that the time-evolution of the density matrix elements can be accurately described using it. Thus, we can exploit the $n$th iterative basis and define a master equation up to the quadratic order in the system--environment coupling and to the first order in $\alpha_{n}(t) = \hbar ||\hat{w}_{n}(t)||/\omega_{01}^{(n)}(t)$, where $E_e^{(n)}-E_g^{(n)} = \hbar \omega_{01}^{(n)}$ such that $E_e^{(n)} = \braket{1|\hat{\tilde{H}}_S^{(n)}|1}$ and $E_g^{(n)} = \braket{0|\hat{\tilde{H}}_S^{(n)}|0}$, as
\begin{widetext}
\begin{equation}
\begin{split}
\dot{\rho}_{gg}^{(n)}   &= -2\Im \mathrm{m}((w_{ge}^{(n)})^*\rho_{ge}^{(n)}) + S(\omega_{01}^{(n)})|m_2^{(n)}|^2 - [S(-\omega_{01}^{(n)})+S(\omega_{01}^{(n)})]|m_2^{(n)}|^2\rho_{gg}^{(n)} + 2[\Im \mathrm{m}(m_2^{(n)})\Im \mathrm{m}(\rho_{ge}^{(n)}) \\ &+\Re \mathrm{e}(m_2^{(n)})\Re \mathrm{e}(\rho_{ge}^{(n)})]S(0)m_1^{(n)} - 2\frac{2S(0)-S(-\omega_{01}^{(n)})-S(\omega_{01}^{(n)})}{\omega_{01}^{(n)}} \{ [\Im \mathrm{m}(m_2^{(n)})\Im \mathrm{m}(w_{ge}^{(n)})+\Re \mathrm{e}(m_2^{(n)})\Re \mathrm{e}(w_{ge}^{(n)})] \\ &\times [\Im \mathrm{m}(m_2^{(n)})\Im \mathrm{m}(\rho_{ge}^{(n)})+\Re \mathrm{e}(m_2^{(n)})\Re \mathrm{e}(\rho_{ge}^{(n)})] \} + 2\frac{2S(0)-S(-\omega_{01}^{(n)})-S(\omega_{01}^{(n)})}{\omega_{01}^{(n)}} \{\Im \mathrm{m}(m_2^{(n)})\Im \mathrm{m}(w_{ge}^{(n)}) \\ &+\Re \mathrm{e}(m_2^{(n)})\Re \mathrm{e}(w_{ge}^{(n)}) \} m_1^{(n)}\rho_{gg}^{(n)} -2\frac{S(0)-S(\omega_{01}^{(n)})}{\omega_{01}^{(n)}}m_1^{(n)}\{ \Im \mathrm{m}(m_2^{(n)})\Im \mathrm{m}(w_{ge}^{(n)})+\Re \mathrm{e}(m_2^{(n)})\Re \mathrm{e}(w_{ge}^{(n)}) \},
\end{split}
\label{eq:master_gg_complete}
\end{equation}
\end{widetext}
and
\begin{widetext}
\begin{equation}
\begin{split}
\dot{\rho}_{ge}^{(n)} &=  \ iw_{ge}^{(n)}(2\rho_{gg}^{(n)}-1)+i(w_{ee}^{(n)}-w_{gg}^{(n)})\rho_{ge}^{(n)}+i\omega_{01}^{(n)}\rho_{ge}^{(n)}-S(\omega_{01}^{(n)})m_1^{(n)}m_2^{(n)} +[S(-\omega_{01}^{(n)})+S(\omega_{01}^{(n)})]m_1^{(n)}m_2^{(n)}\rho_{gg}^{(n)} \\ &-2S(0)(m_1^{(n)})^2\rho_{ge}^{(n)} - i[S(-\omega_{01}^{(n)})+S(\omega_{01}^{(n)})]m_2^{(n)}[\Im \mathrm{m}(\rho_{ge}^{(n)})\Re \mathrm{e}(m_2^{(n)})-\Im (m_2^{(n)})\Re \mathrm{e}(\rho_{ge}^{(n)})] \\ &- 2\frac{2S(0)-S(-\omega_{01}^{(n)})-S(\omega_{01}^{(n)})}{\omega_{01}^{(n)}} (m_1^{(n)})^2w_{ge}^{(n)}\rho_{gg}^{(n)}  + 2\frac{S(0)-S(\omega_{01}^{(n)})}{\omega_{01}^{(n)}}(m_1^{(n)})^2w_{ge}^{(n)} \\ &- im_2^{(n)}\frac{S(-\omega_{01}^{(n)})-S(\omega_{01}^{(n)})}{\omega_{01}^{(n)}}\{\Im \mathrm{m}(m_2^{(n)})\Re \mathrm{e}(w_{ge}^{(n)})-\Im \mathrm{m}(w_{ge}^{(n)})\Re \mathrm{e}(m_2^{(n)})\} \\ &- 2\frac{2S(0)-S(-\omega_{01}^{(n)})-S(\omega_{01}^{(n)})}{\omega_{01}^{(n)}}m_1^{(n)} \{im_2^{(n)}[\Im \mathrm{m}(w_{ge}^{(n)})\Re \mathrm{e}(\rho_{ge}^{(n)})-\Im \mathrm{m}(\rho_{ge}^{(n)})\Re \mathrm{e}(w_{ge}^{(n)})] \\ &- [\Im \mathrm{m}(m_2^{(n)})\Im \mathrm{m}(w_{ge}^{(n)})+\Re \mathrm{e}(m_2^{(n)})\Re \mathrm{e}(w_{ge}^{(n)})]\rho_{ge}^{(n)}\}.
\end{split}
\label{eq:master_ge_complete}
\end{equation}
\end{widetext}
Furthermore, we denote $m_1^{(n)} = \braket{g^{(n)}|\hat{A}|g^{(n)}}$, $m_2^{(n)} = \braket{g^{(n)}|\hat{A}|e^{(n)}}$, $w_{gg}^{(n)} = -i\braket{0|\hat{D}_{n}^{\dagger}\dot{\hat{D}}_{n}|0}$, $w_{ee}^{(n)} = -i\braket{1|\hat{D}_{n}^{\dagger}\dot{\hat{D}}_{n}|1}$ and $w_{ge}^{(n)} = -i\braket{0|\hat{D}_{n}^{\dagger}\dot{\hat{D}}_{n}|1}$. The reduced spectral density of the noise source is defined as $S(\omega) = \int_{-\infty}^{\infty} d\tau \mathrm{Tr}_E \{\hat{\rho}_E \hat{X}(\tau) \hat{X}(0)\} e^{i\omega \tau}/\hbar^2$. Similarly to Refs.~[\onlinecite{prl105/030401, prb82/134517, pra82/062112}], we assume that the system is in the Markov regime, the system time scales are longer than the environment autocorrelation time leading to neglecting the Lamb shift, and the approximation of adiabatic rates applies. These assumptions and the time scale separation they lead to are described in detail in Ref.~\onlinecite{pra82/062112}.

As described above, the benefit of defining the successive coordinate transformations of the Hamiltonian is that the corresponding master equation is up to the first order in $\alpha_{n}(t)$, thus describing the evolution of the system more accurately, in the restricted sense, as $n$ increases. Defining a master equation of arbitrary order in $\alpha_1(t)$ using the original methods \cite{prl105/030401, prb82/134517, pra82/062112} is possible, but the effort required renders such derivations highly unpractical.

It has been shown \cite{prl105/030401, prb82/134517} that assuming a zero-temperature environment and taking the quasi-stationary limit, the master equation in the lowest order leads to $\rho_{gg}^{(1)} = 1 + O(\alpha_1^2)$ and $\rho_{ge}^{(1)} = -w_{ge}^{(1)}/\omega_{01}^{(1)} + O(\alpha_1^2)$. This translates to $\rho_{gg}^{(2)} = 1 + O(\alpha_1^2)$ and $\rho_{ge}^{(2)} = 0 + O(\alpha_1^2)$ showing that in the first order in $\alpha_1$, the density matrix $\hat{\tilde{\rho}}^{(2)}$ describes the evolution of a pure state. This is a remarkable result validating that the master equation in Refs.~[\onlinecite{prl105/030401, prb82/134517, pra82/062112}] ensures relaxation to $\ket{g^{(2)}}$ up to the first order in $\alpha_1$. Similarly, using our master equation for $\hat{\tilde{\rho}}^{(n)}$ ensures that the relaxation takes the system to $\ket{g^{(n+1)}}$ up to the first order in $\alpha_{n}$. Especially, in the limit $n\rightarrow \infty$, the rotational terms $w_{kl}^{(n)}$, $k,l \in \{g,e\}$, in the master equation vanish and the basis $\{\ket{g^{(n)}}, \ket{e^{(n)}}\}|_{n\rightarrow \infty}$ fully describes the steering assuming that the process of basis rotations converges. The requirements for the convergence or the number of transformations up to which the iterative procedure suppresses the time-dependence of the Hamiltonian of the system \cite{prsla392/45} are not studied in this paper, as it turns out in Sec.~\ref{sec:sluice} that a small number of transformations allows one to capture the main effect of this scheme.

\section{\label{sec:sluice}Cooper pair sluice}

\subsection{\label{subsec:sluice_ana}Definitions}

We introduce the Cooper pair sluice \cite{prl91/177003} as a physical realization of a steered two-level system. The charge pumped through the sluice establishes a connection to geometric phases \cite{prb60/R9931, prb68/020502(R), prb73/214523, prl100/117001, prl100/177201} acquired during the adiabatic evolution and provides a physical observable. We aim to study and improve on the recent theoretical pumping results \cite{prl105/030401, prb82/134517} using the high-order effective theory.

\begin{figure*}
\includegraphics[width=15cm]{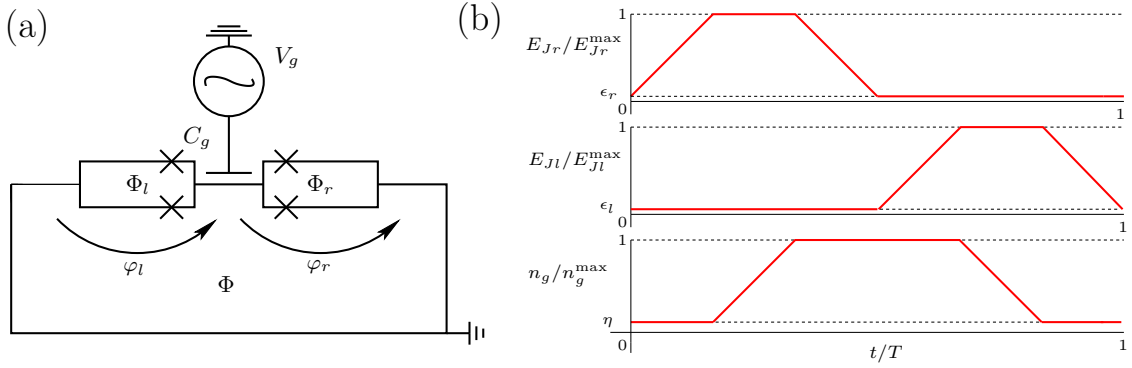}
\caption{\label{fig:merged}(Color online) (a) Circuit diagram of the Cooper pair sluice. The fluxes threading the left and right SQUIDs are denoted by $\Phi_l$ and $\Phi_r$, and $\Phi$ denotes the total flux threading the large superconducting loop. The phase differences over the SQUIDs are marked by $\varphi_l$ and $\varphi_r$ and they are defined in the direction specified by the arrows in the figure. The gate capacitance $C_g$ is used to manipulate the island charge with the gate voltage $V_g$. (b) The time dependence of the control parameters $E_{Jl}$, $E_{Jr}$, and $n_g$ during a pumping cycle. The residual values $\epsilon_l$, $\epsilon_r$, and $\eta$ allow for non-ideal SQUIDs and gate control.}
\end{figure*}

The Cooper pair sluice shown in Fig.~\ref{fig:merged}(a) is comprised of a superconducting island separated by two SQUIDs \cite{ItS}, each involving two Josephson junctions.
If we assume that the self-inductances of the SQUID loops are negligible, the SQUIDs operate as tunable Josephson junctions whose Josephson energies $E_{Jk}(\Phi_k)$, where $k \in \{l,r\}$, are determined by the external fluxes threading the loops.  The sluice Hamiltonian is
\begin{equation}
\hat{H}_S = E_C(\hat{n}-n_g)^2 - E_{Jr}\cos \left(\hat{\phi}+\frac{\varphi}{2}\right) - E_{Jl}\cos \left(\hat{\phi}-\frac{\varphi}{2}\right),
\label{eq:H_sluice}
\end{equation}
where the Coulomb energy for one excess Cooper pair is $E_C = 2e^2/C_{\Sigma}$ and the gate charge is defined in units of $2e$ as $n_g = V_gC_g/(2e)$. Here $C_g$ is the gate capacitance and $C_{\Sigma}$ stands for the total capacitance of the island. The operator describing the phase on the island is $\hat{\phi} = (\hat{\varphi}_r - \hat{\varphi}_l)/2$ and its canonical conjugate, the operator describing the number of excess Cooper pairs on the island, is $\hat{n} = -i\partial_{\hat{\phi}}$. The gauge-invariant phase difference over the device $\varphi = \hat{\varphi}_r + \hat{\varphi}_l$ is determined by $\varphi = 2\pi \Phi / \Phi_0$ and kept constant during the evolution. 

We denote $\hat{n}_k = -i\partial_{\hat{\varphi}_k}$ ($k \in \{l,r\}$) as the Cooper pair number operator of the $k$th SQUID and write the average value of the current through the $k$th SQUID as \cite{prb82/134517}
\begin{equation}
\braket{\hat{I}_k} = \frac{2ei}{\hbar} \left( \mathrm{Tr}_S\{ \hat{\rho}_S[\hat{n}_k,\hat{H}_S]\} + \mathrm{Tr}\{ \hat{\rho}[\hat{n}_k,\hat{V}]\} \right),
\label{eq:current_ave}
\end{equation}
where $\hat{I}_k$ denotes the respective current operator. If the environment does not directly induce a current, that is, $[\hat{n}_k,\hat{V}]=0$, Eq.~(\ref{eq:current_ave}) defines the usual current operator \cite{prb73/214523}
\begin{equation}
\hat{I}_k = -\frac{2ei}{\hbar}[\hat{H}_S,\hat{n}_k] = \frac{2e}{\hbar} \frac{\partial \hat{H}_S}{\partial \hat{\varphi}_k}.
\label{eq:current_operator}
\end{equation}

If we set $E_C \gg \text{max}\{E_{Jl}, E_{Jr}\}$ and $n_g \approx 1/2$, the dynamics are accurately described by the two lowest charge states allowing us to apply the preceding two-state theory. We denote $\ket{0}$ and $\ket{1}$ as the states with no and with one excess Cooper pair on the island defining our fixed basis. We study the capacitive coupling of the environment to the system by introducing voltage fluctuations $\delta\hat{V}_g(t)$ at the gate of the sluice \cite{prl105/030401, prb82/134517}. The coupling operator then becomes $\hat{V}_g = -eg\hat{\sigma}_z \otimes \delta \hat{V}_g(t)$ where $\hat{\sigma}_z = \ket{0}\bra{0}-\ket{1}\bra{1}$ and $g = C_g/C_{\Sigma}$ denotes the strength of the coupling. The coupling operator has been selected traceless in the two-state basis by adding an operator comparable to the identity operator to adopt a convention used in the derivation of the master equation \cite{pra82/062112}. Such a selection can be applied to any coupling operator and it does not reduce the generality of the master equation. Since $[\hat{n}_k,\hat{V}]=0$, Eqs.~(\ref{eq:H_sluice}) and (\ref{eq:current_operator}) imply that $\hat{I}_k = 2eE_{Jk} \sin(\hat{\varphi}_k)/\hbar$.

Assume that the noise source is a resistor in thermal equilibrium; a situation which can be engineered in the physical realization of the sluice. If we consider the voltage noise of a resistor grounded at one end and connected to the gate by a low impedance circuit at the other end, the reduced spectral density of the noise source at the gate becomes \cite{rmp82/1155} $S(\omega) = 2R_g\omega/[\hbar(1-e^{-\hbar\omega/(k_BT_R)})]$, where $R_g$ is the effective resistance of the noise source and $T_R$ is the resistor temperature. We note that the detailed balance condition $S(\omega)=e^{\hbar\omega/(k_BT_R)}S(-\omega)$ applies. Furthermore, we introduce dephasing to the system by assuming that $S(0) = 2k_BT_0R_g/\hbar^2$, where $T_0$ is the effective dephasing temperature.

We denote the matrix elements of the current operator of the $k$th SQUID by $I_{k,rs}^{(n)}=\braket{r^{(n)}|\hat{I}_k|s^{(n)}}$. The expectation value of the current using the adiabatic basis is
\begin{equation}
\begin{split}
\braket{\hat{I}_k} = \rho^{(1)}_{gg}I_{k,gg}^{(1)} + \rho_{ee}^{(1)}I_{k,ee}^{(1)} + 2\Re \mathrm{e}(\rho_{ge}^{(1)}I_{k,eg}^{(1)}),
\end{split}
\label{eq:<I>}
\end{equation}
since $\hat{I}_k$ is Hermitian. The first two terms are the dynamic supercurrent through the junction and the third term describes the geometric part of the current \cite{prb73/214523, prl100/177201, prl105/030401, prb82/134517}. The pumped charge corresponding to the geometric contribution becomes
\begin{equation}
\begin{split}
Q_{G,k} = 2\Re \mathrm{e}\bigg[ \int_t^{t+T} \, dt' \rho_{ge}^{(1)}(t') I_{k,eg}^{(1)}(t') \bigg],
\end{split}
\label{eq:<Q>}
\end{equation}
where $T$ is the length of the closed cycle. 

The definition for the different terms in the average current only applies directly for the adiabatic basis. This implies that if we pursue to obtain the geometric current using higher order bases, $\rho_{ge}^{(1)}$ should be written using the density matrix elements in the basis where the evolution takes place in our calculations. The adiabatic density matrix element can be written as
\begin{equation}
\begin{split}
\rho_{ge}^{(1)} &= \braket{g^{(1)}|\hat{\rho}_S|e^{(1)}} \\ &= \braket{0|\hat{D}_1^{\dagger}\hat{\rho}_S\hat{D}_1|1} = \braket{0|\hat{D}_S^{(n)}\hat{\tilde{\rho}}_S^{(n)}(\hat{D}_S^{(n)})^{\dagger}|1}.
\end{split}
\label{eq:adi_rho_succ}
\end{equation}
Using this, we can rewrite the integrand in the pumped charge
\begin{equation}
\begin{split}
& \Re \mathrm{e}\{\rho_{ge}^{(1)}I_{k,eg}^{(1)}\} \\ &= \rho_{gg}^{(n)} \Re \mathrm{e}\{ I_{k,eg}^{(1)}(\braket{0|\hat{D}_S^{(n)}|0}\braket{0|(\hat{D}_S^{(n)})^{\dagger}|1} \\ &-\braket{0|\hat{D}_S^{(n)}|1}\braket{1|(\hat{D}_S^{(n)})^{\dagger}|1}) \} \\ &+ \Re \mathrm{e}\{\rho_{ge}^{(n)}\} \Re \mathrm{e}\{ I_{k,eg}^{(1)}(\braket{0|\hat{D}_S^{(n)}|0}\braket{1|(\hat{D}_S^{(n)})^{\dagger}|1} \\ &+\braket{0|\hat{D}_S^{(n)}|1}\braket{0|(\hat{D}_S^{(n)})^{\dagger}|1}) \} \\ &- \Im \mathrm{m}\{\rho_{ge}^{(n)}\} \Im \mathrm{m}\{ I_{k,eg}^{(1)}(\braket{0|\hat{D}_S^{(n)}|0}\braket{1|(\hat{D}_S^{(n)})^{\dagger}|1} \\ &-\braket{0|\hat{D}_S^{(n)}|1}\braket{0|(\hat{D}_S^{(n)})^{\dagger}|1}) \} \\ &+ \Re \mathrm{e}\{ I_{k,eg}^{(1)}\braket{0|\hat{D}_S^{(n)}|1}\braket{1|(\hat{D}_S^{(n)})^{\dagger}|1} \}.
\end{split}
\label{eq:rerhoI_succ}
\end{equation}
Equation~(\ref{eq:rerhoI_succ}) enables us to calculate the pumped charge when the time-evolution of the $n$th basis density matrix is known. Note that for the adiabatic basis, Eq.~(\ref{eq:rerhoI_succ}) reduces to the form corresponding to Eq.~(\ref{eq:<Q>}).

\subsection{\label{subsec:sluice_num}Effect of the environment on the pumped charge}

We use the parameter cycle shown in Fig.~\ref{fig:merged}(b) for the pumping and ensure the smoothness of the parameter functions in time using trigonometric interpolation dividing the total cycle time into 201 equidistant points. This is a necessary step for the simulations as exploiting the high-order bases requires that the high-order temporal derivatives of the parameter functions stemming from $\hat{w}_n$ are non-divergent. The dynamics of the quantum system are solved numerically from Eqs.~(\ref{eq:master_gg_complete}) and (\ref{eq:master_ge_complete}) utilizing the effective Hamiltonians introduced in Sec.~\ref{sec:model}. The density matrix and any physical observables are recorded in the steady state, that is, after sufficiently many cycles such that the system evolution of consecutive cycles is identical.

We begin by studying the effect of using the higher-order bases when describing the dynamics of the sluice in a zero-temperature environment. Using the lowest-order approximation $n=1$, the analytical result of the environment inducing ground-state evolution in the adiabatic limit has been demonstrated using numerical simulations of the pumped charge \cite{prl105/030401, prb82/134517}. Additionally, increasing the pumping frequency has been shown to induce regions where either the nonadiabatic transitions or relaxation dominates depending on the ratio $\alpha_1/g$. However, it turns out that a region in the $(\alpha_1,g)$-space emerges where the pumped charge is unphysically overestimated. This is a direct result of the master equation not strictly ensuring the positivity of the density matrix in any finite order, that is, it does not reduce to the standard Lindblad form.

\begin{figure*}
\centering
\includegraphics[width=15cm]{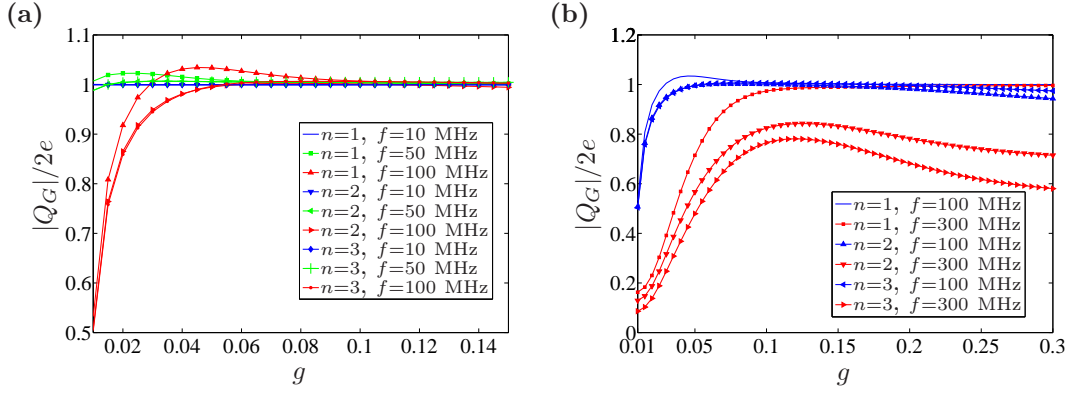}
\caption{\label{fig:Qp_g_10_50_100_T0}(Color online) Charge pumped through the Cooper pair sluice during one cycle assuming that the temperature of the environment is $T_R=0$ (a) with the pumping frequency $f=10$, 50, and 100 MHz and (b) $f=100$ and 300 MHz. The simulations are carried out setting the number of coordinate transformations to $n=1$, 2, and 3. The physical parameters used in the simulations are $T_0=0.1$ K, $R_g=300$ k$\Omega$, $E_C/k_B=1$ K, $E_{Jr}^{\mathrm{max}}/k_B = E_{Jl}^{\mathrm{max}}/k_B =0.1$ K, $\epsilon_r=\epsilon_l=0.03$, $n_g^{\mathrm{max}}=0.8$, $\eta=0.25$ and $\varphi=\pi/2$.}
\end{figure*}

We present the pumped charge using the basis $\{\ket{g^{(n)}}, \ket{e^{(n)}}\}$ with $n=1,2,3$ as a function of the coupling strength $g$ for different pumping frequencies in Fig.~\ref{fig:Qp_g_10_50_100_T0}(a) assuming a zero-temperature environment. We explore the regime where the strength of the environmental coupling is small to ensure that we remain close to the weak coupling limit.
In the adiabatic region ($f=10$ MHz), all orders of approximation indicate ground-state pumping for all environmental coupling strengths as predicted in Sec.~\ref{sec:master}. By increasing the pumping frequency, we observe the emergence of the two pumping regimes mentioned above. Additionally, Figure~\ref{fig:Qp_g_10_50_100_T0}(a) illustrates how increasing the coupling strength does not lead to ground-state pumping beyond the adiabatic region for $n>1$. The reason for this phenomenon stems from the structure of the superadiabatic bases. For nonadiabatic evolution, increasing the coupling strength leads to relaxation to $\ket{g^{(n+1)}}$ up to the first order in $\alpha_{n}$ which translates to ideal pumping only for $n=1$ as the system is forced to the solution of the adiabatic limit, from which the asymptotic solutions for $n>1$ generally deviate in all orders of the local adiabatic parameter. The adiabatic solution for the pumped charge assuming similar SQUIDS $\epsilon = \epsilon_r = \epsilon_l$ has been derived previously \cite{prb73/214523} as $Q_G = 2e(1-2\epsilon\cos\varphi)$. To make the discrepancy between the high-order bases and the case $n=1$ more visible, we present the pumped charge up to high coupling strengths in Fig.~\ref{fig:Qp_g_10_50_100_T0}(b). The regime of increased coupling strength is beyond the range of validity of our approach but shows how the master equation properly displays the distinctions between the different bases as they approach the asymptotic relaxation dominated solutions.

\begin{figure*}
\centering
\includegraphics[width=15cm]{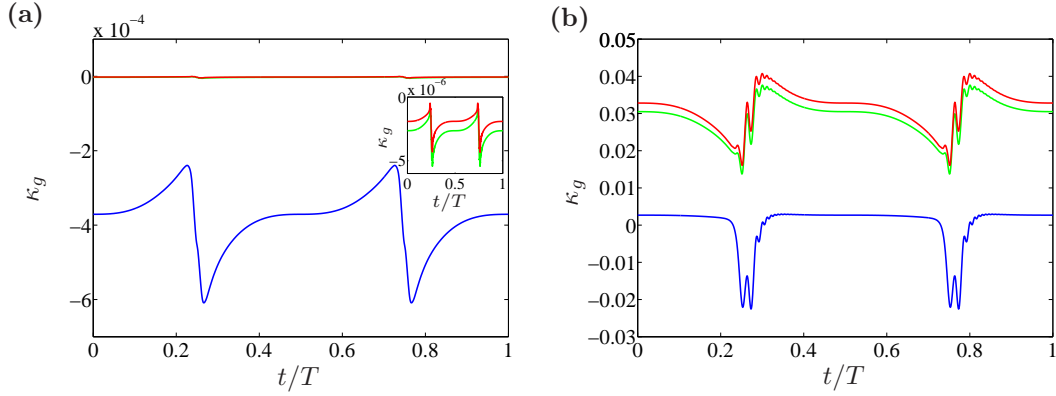}
\caption{\label{fig:kappa_10_100}(Color online) Lower eigenvalue $\kappa_g$ of the reduced density operator after $n=1$, 2, and 3 coordinate transformations from bottom to top (a) for $f=10$ MHz and $g=0.01$ and (b) for $f=100$ MHz and $g=0.03$. Inset of (a) shows $\kappa_g$ for $f=10$ MHz and $n=2$ and 3 from bottom to top. Other physical parameters are identical to those used in Fig.~\ref{fig:Qp_g_10_50_100_T0}.}
\end{figure*}

From Fig.~\ref{fig:Qp_g_10_50_100_T0}, it is clear that the charge overestimation observed for $n=1$ is alleviated by utilizing the high-order bases. As the high-order bases follow the exact evolving closed system state more closely, the non-adiabatic transitions disturb the mixed state less. The major contribution in alleviating the lack of complete positivity is already given by $n=2$ and the subsequent third basis rotation has little effect on the pumped charge in comparison. The effect of using the higher-order bases is evident from studying not only a set of observables but from the density matrix itself. We present the lower eigenvalue $\kappa_g$ of the steady state density operator $\hat{\tilde{\rho}}_S^{(n)}$ in Fig.~\ref{fig:kappa_10_100}.
The non-positivity is greatly reduced and for some parameter values, completely removed by doing one or more further rotations beyond the adiabatic basis.

\subsection{\label{subsec:sluice_num2}Pumped current and the breakdown of adiabaticity}

So far, experimental results for Cooper pair pumping using the sluice have been scarce \cite{prb71/012513, apl90/082102, prl100/177201}. However, the breakdown of adiabaticity has been observed by studying the pumped current $I_G$ as a function of the pumping amplitude $n_{\mathrm{range}} = \text{max}\{n_g\}-\text{min}\{n_g\}$, that is, at high pumping amplitudes the pumped current has been noticed to deviate from the analytical result in the adiabatic limit $|I_G|=2en_{\mathrm{\scriptscriptstyle CP}}f$, where $n_{\mathrm{\scriptscriptstyle CP}}$ is the number of Cooper pairs transported ideally through the sluice per cycle \cite{prl100/177201}. The number of transported Cooper pairs can be experimentally dictated by adjusting the gate voltage, more spesifically, by altering $n_{\text{range}}$ so that the ideally transported average current corresponds to $|I_G|=2en_{\mathrm{range}}f$. Unfortunately, our master equation cannot be directly used to simulate the effect of altering the pumping amplitude since it is defined in a two-state basis which requires that $n_g$ remains approximately half-integer during the evolution. Any selection of the two charge states as the fixed basis enables the maximum pumping amplitude of one Cooper pair per cycle.

\begin{figure*}
\centering
\includegraphics[width=15cm]{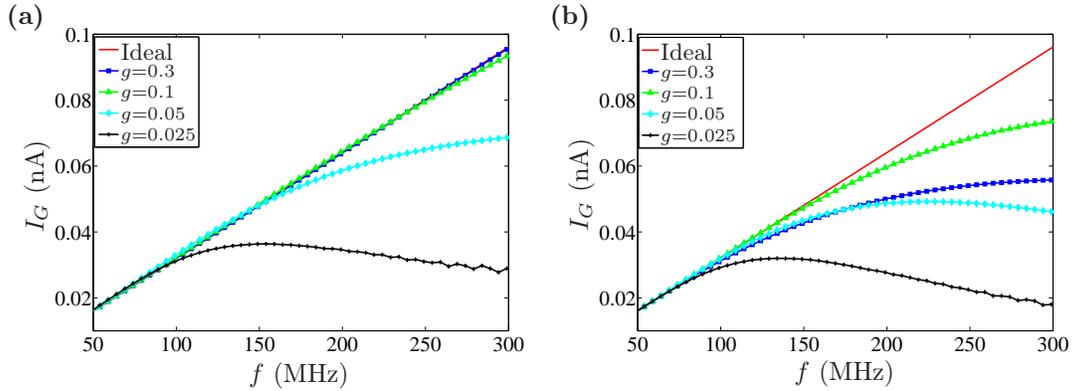}
\caption{\label{fig:IG}(Color online) Current pumped through the sluice during one cycle as a function of the pumping frequency $f$ with coupling strengths $g=0.025$, 0.05, 0.1 and 0.3 using (a) $n=$1 and (b) $n=3$. Other physical parameters are identical to those used in Fig.~\ref{fig:Qp_g_10_50_100_T0}. The linear solid line is the ideal pumping curve for the two-state model $|I_G|=2ef$.}
\end{figure*}

Even though we cannot modify the pumping amplitude, we can still simulate the breakdown of the adiabaticity by altering the pumping frequency. If we assume that the deviation from the adiabatic behaviour in the experiments is due to increase in the pumping speed caused by the amplitude growth using a constant cycle time, the effect of decreasing the total cycle time should be similar. We choose a frequency range beyond the strict adiabatic limit to model experiments carried out with finite cycle times. The pumped current is shown in Fig.~\ref{fig:IG}(a) using $n=1$ and in Fig.~\ref{fig:IG}(b) using $n=3$.
The behaviour of the pumped current with high frequencies is only suggestive but clearly indicates that the effect of the number of coordinate transformations increases with the pumping frequency. The physically most relevant features are found near the point where the adiabaticity breaks down. As we utilize a more accurate description of the dynamics, the point where the adiabaticity of the system is broken becomes more robust against changes in the environment. Furthermore, we observe the emergence of an optimal coupling strength, with which the ideal ground-state pumping is conserved up to the highest frequency. This feature could have also been anticipated from Fig.~\ref{fig:Qp_g_10_50_100_T0} where we observe the emergence of a maximum pumped charge as a function of the coupling srength for any given frequency far from the adiabatic limit if $n>1$. This corresponds to the coupling strength, below which the non-adiabatic transitions reduce the pumped charge and above which the relaxation takes the system away from the solution in the adiabatic limit. The optimal coupling strength can be probed and exploited utilizing the environment engineering scheme presented in Ref.~\onlinecite{prb82/134517}.

\begin{figure*}
\centering
\includegraphics[width=15cm]{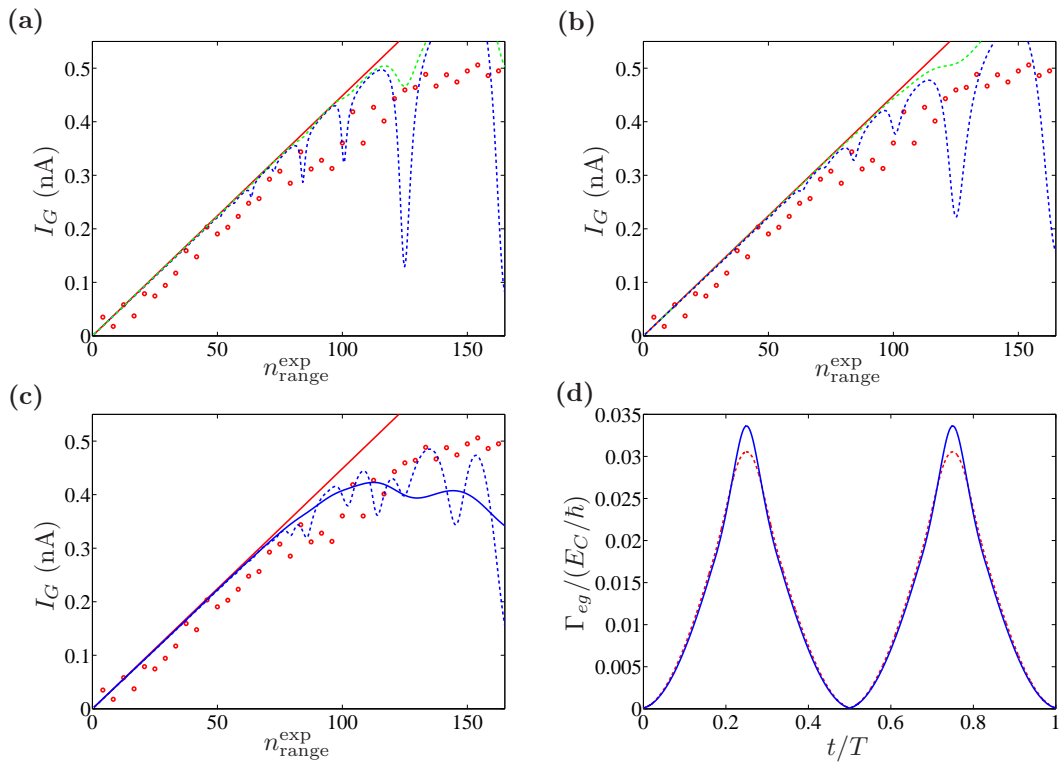}
\caption{\label{fig:expIG}(Color online) (a)-(c) Pumped current per cycle given by the experiment (circles) and by the simulation (dashed lines). The linear solid line is the ideal pumping curve assuming that the number of geometrically transported Cooper pairs is given by the pumping amplitude $|I_G|=2en_{\mathrm{range}}^{\mathrm{exp}}f_{\mathrm{exp}}$. The parameters from the experiments are $E_C/k_B=2$ K, $E_{Jr}^{\mathrm{max}}/k_B = E_{Jl}^{\mathrm{max}}/k_B =0.7151$ K, $\epsilon_r = \epsilon_l =0.05$, $\varphi=\pi/2$, and $f_{\mathrm{exp}}=14$ MHz. For the simulations, we use $T_R=200$ mK, $R_g=300$ k$\Omega$, $n_g^{\mathrm{max}}=0.8$, and $\eta=0.25$ giving $n_{\mathrm{range}}^{\mathrm{sim}}=0.6$. The effect of dephasing is studied by using (a) $T_0=0$ K and (b) $T_0=0.1$ K, where the coupling strength is $g=0.025$, 0.05 from bottom to top. In (c), we present the pumped current for $T_0=1$ K and $g=0.025$ (solid), and the averaged pumped current for $T_0=0.1$ K and $g=0.025$ (dashed). The corresponding relaxation rate of the system (solid), and the averaged relaxation rate (dashed) are given in (d) as a function of time. Experimental data courtesy of M. M\"ott\"onen, J. J. Vartiainen, and J. P. Pekola.}
\end{figure*}

We turn our attention to modeling the breakdown characteristics of the actual experimentally pumped current in Ref.~\onlinecite{prl100/177201}. The simulation is motivated by the above discussion on the pumping speed and we establish an equality between the experimental pumping speed $n_{\mathrm{range}}^{\mathrm{exp}} \times f_{\mathrm{exp}}$, where $f_{\mathrm{exp}}$ is maintained at a constant value, and the pumping speed in our simulation $n_{\mathrm{range}}^{\textrm{sim}} \times f$, where $n_{\mathrm{range}}^{\mathrm{sim}}$ is a constant to ensure that the two-state approximation holds. A comparison between the experimental results and our simulations is presented in Fig.~\ref{fig:expIG}.

The system parameters used in the simulations are estimated from the experiments and can contain up to 20\% error. To simulate the experiments, the aforementioned equality defines the needed frequency by $f= n_{\mathrm{range}}^{\mathrm{exp}} \times 23.33$ MHz so that the point where the adiabaticity breaks down according to the experimental data would require a simulation frequency of approximately 1.1 GHz. This is significantly higher than the breakdown frequency in our previous simulations and tests the limits of validity of our approach. 

Assuming both vanishing dephasing in Fig.~\ref{fig:expIG}(a) and non-vanishing dephasing in Fig.~\ref{fig:expIG}(b), the simulated pumped current exhibits strong oscillatory behavior allowing for weak predictability of the exact breakdown characteristics. The observed behavior is due to quantum interference between driving-induced excitations generated at different times. The instantaneous transition probability is not only dependent on the energy gap but also on the phases accumulated during the quantum evolution. Especially, if the phase difference between two successive excitations in time is a multiple of $2\pi$, the transition probability reaches its maximum due to constructive interference and the geometric current obtains a downward resonance peak. Studying the phase accumulation during the quantum evolution allows us to estimate the resonance peak positions and compare them with the simulated pumped current.

The time-evolution of the energy gap of the system is symmetric with respect to the mid-point of each pumping cycle corresponding to times $t_{\mathrm{mid}}^{(i)}=(2i+1) \times T/2$, $i \in \{0,1,2\dots\}$. In addition, the energy gap decreases during the two gate manipulations and as both manipulations in Figs.~\ref{fig:expIG}(a) and \ref{fig:expIG}(b) are symmetric with respect to the degeneracy point $n_g = 1/2$, the energy gap reaches its minima at points we denote as $t_i=(2i+1) \times T/4$. Hence, it is enough to study the phase accumulation between different $t_i$ to depict the resonance behavior. 

In adiabatic evolution, the number of coordinate transformations we perform has a profound effect on the observed accumulated phase, that is, each transformation takes us to a new basis in which the phase accumulates differently. More exactly, the closed system state $\ket{\Psi(t)}$ follows the Schr\"odinger equation $i\hbar\ket{\dot{\Psi}(t)} = \hat{H}_S(t)\ket{\Psi(t)}$ at all times. After $m$ transformations, the state $\ket{\Psi^{(m)}(t)} = \hat{D}_1(t)\hat{D}_S^{(m)}(t)\ket{\Psi(t)}$ follows the transformed equation $i\hbar\ket{\dot{\Psi}^{(m)}(t)} = (\hat{\tilde{H}}_S^{(m)}(t)+\hbar \hat{w}_m(t))\ket{\Psi^{(m)}(t)}$. Assuming adiabaticity in this basis, that is, the exact evolving state remains in the $k$th eigenstate of $\hat{\tilde{H}}_S^{(m)}(t)+\hbar \hat{w}_m(t)$, the $m$-times transformed evolving state can be written, similarly to Ref.~\onlinecite{ajp66/431}, as
\begin{equation}
\begin{split}
\ket{\Psi^{(m)}(t)} = \exp \left( i\alpha_k^{(m)}(t) - \frac{1}{\hbar} \int_{t_{\mathrm{in}}}^{t} d\tau \tilde{E}_k^{(m)}(\tau) \right) \ket{\tilde{k}^{(m)}(t)},
\end{split}
\label{eq:respre1}
\end{equation}
where we assumed that $\ket{\Psi^{(m)}(t_{\mathrm{in}})} = \ket{\tilde{k}^{(m)}(t_{\mathrm{in}})}$, $(\hat{\tilde{H}}^{(m)}_S(t) + \hbar \hat{w}_m(t))\ket{\tilde{k}^{(m)}(t)} = \tilde{E}_k^{(m)}(t)\ket{\tilde{k}^{(m)}(t)}$, and $\alpha_k^{(m)}(t)$ describes the geometric phase contribution. Using the familiar notation, this implies $\tilde{E}_k^{(m)}(t) = E_k^{(m+1)}(t)$ and $\ket{\tilde{k}^{(m)}(t)} = \hat{D}_{m+1}(t)\ket{k_f}$. The accumulated quantum phases can be obtained from Eq.~(\ref{eq:respre1}) following the derivation in Ref.~\onlinecite{ajp66/431}. However, in our dissipative calculations, after $n$ transformations, we take $\hbar \hat{w}_n + \hat{\tilde{V}}^{(n)}$ as the perturbation. This means that for the dissipative simulations, the relevant reference frame after $n$ transformations is given by the eigenbasis of $\hat{H}_S(t)$ for $n=1$ and $\hat{\tilde{H}}^{(n-1)}_S(t) + \hbar \hat{w}_{n-1}(t)$ for $n>1$, that is, $\tilde{E}_k^{(n-1)}(t) = E_k^{(n)}(t)$ and $\ket{\tilde{k}^{(n-1)}(t)} = \hat{D}_{n}(t)\ket{k_f}$.

Concentrating on the two-state model, we present the accumulated phases after $n$ transformations. We assume that the transitions taking place at $t_i$ are instantaneous and the system evolves adiabatically between them. The dynamically accumulated phase obtained during adiabatic evolution between $t_i$ and $t_{i+1}$ for the eigenstates in the relevant reference frame after $n$ transformations is given by $\Theta_{D,k}^{(n)}(t_i,t_{i+1}) = -\int_{t_i}^{t_{i+1}} dt E^{(n)}_k/\hbar$, $k \in \{g,e\}$, and it is equal for any successive two points due to symmetry. The geometrically accumulated phase $\Theta_{G,k}^{(n)}(t_i,t_{i+1})$ is not the usual Berry phase since $t_i$ with odd and even $i$ describe different points in the Hamiltonian space and, thus, the path traversed between $t_i$ and $t_{i+1}$ is not closed. Assuming adiabatic evolution, we can write the noncylic geometric phase acquired by the $k$th eigenstate between two successive transitions using Eq.~(\ref{eq:respre1}) as \cite{ajp66/431, pra82/062112}
\begin{equation}
\begin{split}
\Theta_{G,k}^{(n)}(t_i,t_{i+1}) &= \arg \{ \braket{\tilde{k}^{(n-1)}(t_i)|\tilde{k}^{(n-1)}(t_{i+1})} \} \\ &+ i \int_{t_i}^{t_{i+1}} dt \braket{\tilde{k}^{(n-1)}(t)|\partial_t|\tilde{k}^{(n-1)}(t)} \\ &= \arg \{ \braket{k_f|\hat{D}_n^{\dagger}(t_i)\hat{D}_n(t_{i+1})|k_f} \} \\ &+ i \int_{t_i}^{t_{i+1}} dt \braket{k_f|\hat{D}_n^{\dagger}(t)\dot{\hat{D}}_n(t)|k_f}.
\end{split}
\label{eq:res1}
\end{equation}
The geometric phase defined in Eq.~(\ref{eq:res1}) is gauge-invariant and, thus, only depends on the traversed path. Using our parameter cycle, the accumulated geometric phase for any successive two points is the same. Note that as the transformations suppress the time-dependence of the effective Hamiltonians, the geometric phase is decreased as $n$ increases. The difference in the total accumulated phase acquired by the eigenstates is
\begin{equation}
\begin{split}
& \Delta \Theta_T^{(n)}(t_i,t_{i+1}) \\ &= \Theta_{T,g}^{(n)}(t_i,t_{i+1}) - \Theta_{T,e}^{(n)}(t_i,t_{i+1}) \\ &= \arg \{ \braket{0|\hat{D}_n^{\dagger}(t_i)\hat{D}_n(t_{i+1})|0} \} - \arg \{ \braket{1|\hat{D}_n^{\dagger}(t_i)\hat{D}_n(t_{i+1})|1} \} \\ & + \int_{t_i}^{t_{i+1}} dt \, [w_{ee}^{(n)}(t)-w_{gg}^{(n)}(t)] + \int_{t_i}^{t_{i+1}} dt \ \omega_{01}^{(n)}(t).
\end{split}
\label{eq:res2}
\end{equation}
For a full cycle, we have $\Delta \Theta_T^{(1)}(t_i,t_{i+2}) \approx 2(\varphi - 2\epsilon\sin\varphi) + \int_{t_i}^{t_{i+2}} \omega_{01}^{(1)}(t)$, where we assume similar SQUIDs $\epsilon = \epsilon_r = \epsilon_l$, as the accumulated geometric phases become the Berry phases in the adiabatic basis \cite{prb73/214523, science318/1889}. 

In addition to the phases accumulated in adiabatic evolution, we must account for any phase shifts occurring at the transition. We apply the adiabatic-impulse model \cite{pr492/1} in the relevant reference frame and describe the non-adiabatic transitions taking place at $t_i$ as instantaneous processes. The model is based on the Landau-Zener approximation \cite{prsla137/696} assuming that in the vicinity of each $t_i$, $\hat{H}_S(t)$ for $n=1$ and $\hat{\tilde{H}}^{(n-1)}_S(t)+\hbar\hat{w}_{n-1}(t)$ for $n>1$ can be linearized in a fixed basis as 
\begin{equation}
\hat{H}_{\mathrm{LZ}}^{(n)}(t) = -\Delta_n/2 \hat{\sigma}_x \mp v_nt/2 \hat{\sigma}_z, 
\label{eq:H_LZ}
\end{equation}
where $\hat{\sigma}_x = \ket{0}\bra{1} + \ket{1}\bra{0}$, the upper sign corresponds to odd $i$ and the lower sign corresponds to even $i$. The representation of the Hamiltonian exploiting a time-independent basis is usually referred to as \textit{diabatic}. The tunneling amplitude $\Delta_n$ and the slope of the linearized energy bias $v_n$ are assumed real. Our Hamiltonians do not exactly linearize into this form, but the off-diagonal elements obtain phase factors related to the phase difference over the sluice for $n=1$ and to the complex phase of $w_{ge}^{(n)}$ for $n>1$. These phase factors can be accounted for by a transformation to a representation where the Hamiltonian linearizes to the Landau-Zener form. The Landau-Zener transition probability from the ground state to the excited state is given by $P_{\mathrm{LZ}}^{(n)} = \exp (-2\pi \delta_n)$, where the adiabaticity parameter is $\delta_n = \Delta_n^2/(4v_n)$.

We can generally linearize the relevant Hamiltonian after $n$ transformations in the diabatic representation in the vicinity of each $t_i$ as
\begin{equation}
\begin{split}
\hat{H}_{\mathrm{lin}}^{(n)}(t) = &-\Delta_n(t_i)e^{i\gamma_n(t_i)}/2 \ket{0}\bra{1} \\ &- \Delta_n(t_i)e^{-i\gamma_n(t_i)}/2 \ket{1}\bra{0} \\ &\mp v_n(t_i)t/2 \hat{\sigma}_z,
\end{split}
\label{eq:H_lin}
\end{equation}
where $\gamma_1(t_i) = \arg \{E_{Jr}(t_i)e^{-i\varphi/2}+E_{Jl}(t_i)e^{i\varphi/2} \}$ and $\gamma_n(t_i) = \arg \{ - w_{ge}^{(n-1)}(t_i) \}$ for $n>1$. Additionally, we have $\Delta_1(t_i) = |E_{Jr}(t_i)e^{-i\varphi/2}+E_{Jl}(t_i)e^{i\varphi/2}|$ and $\Delta_n(t_i) = |2\hbar w_{ge}^{(n-1)}(t_i)|$ for $n>1$. Since the gate charge is altered symmetrically with respect to the degeneracy point $n_g = 1/2$ in Fig.~\ref{fig:expIG}, the linearization yields $v_1(t_i) = -2E_C\dot{n}_g(t_i)$ and $v_n(t_i) = -2[\dot{E}_g^{(n-1)}(t_i))+\hbar \dot{w}_{gg}^{(n-1)}(t_i)]$ for $n>1$. Then the state of the system $\ket{\psi(t)}$ evolves according to $i\hbar \ket{\dot{\psi}(t)} = \hat{H}_{\mathrm{lin}}^{(n)}(t)\ket{\psi(t)}$ near $t_i$. Defining a transformation
\begin{equation}
\begin{split}
\hat{U}_{\mathrm{LZ}}^{(n)}(t) = e^{i\gamma_n(t)/2}\ket{0}\bra{0} + e^{-i\gamma_n(t)/2}\ket{1}\bra{1},
\end{split}
\label{eq:LZtrans}
\end{equation}
yields that the transformed state $\ket{\varphi(t)} = (\hat{U}_{\mathrm{LZ}}^{(n)}(t))^{\dagger}\ket{\psi(t)}$ follows $i\hbar \ket{\dot{\varphi}(t)} = \hat{H}_{\mathrm{LZ}}^{(n)}(t)\ket{\varphi(t)}$ near $t_i$ and we can apply the adiabatic impulse model. The evolution operator describing a single Landau-Zener transition is \cite{pr492/1}
\begin{equation}
\begin{split}
&\hat{N}_{\mathrm{LZ}}^{(n)}(t_i-\Delta t,t_i+\Delta t) \\ &= \sqrt{1-P_{\mathrm{LZ}}^{(n)}} e^{-i\widetilde{\varphi}_{S}^{(n)}}\ket{e_{\mathrm{LZ}}^{(n)}(t_i)}\bra{e_{\mathrm{LZ}}^{(n)}(t_i)} \\ &- \sqrt{P_{\mathrm{LZ}}^{(n)}}\ket{e_{\mathrm{LZ}}^{(n)}(t_i)}\bra{g_{\mathrm{LZ}}^{(n)}(t_i)} \\ &+ \sqrt{P_{\mathrm{LZ}}^{(n)}}\ket{g_{\mathrm{LZ}}^{(n)}(t_i)}\bra{e_{\mathrm{LZ}}^{(n)}(t_i)} \\ &+ \sqrt{1-P_{\mathrm{LZ}}^{(n)}}e^{i\widetilde{\varphi}_{S}^{(n)}}\ket{g_{\mathrm{LZ}}^{(n)}(t_i)}\bra{g_{\mathrm{LZ}}^{(n)}(t_i)},
\end{split}
\label{eq:res3}
\end{equation}
where $\Delta t$ is a short time-step, $\widetilde{\varphi}_{S}^{(n)} = \varphi_{S}^{(n)} - \pi/2$, $\varphi_{S}^{(n)} = \pi/2 + \delta_n(\ln \delta_n - 1) + \arg \{ \Gamma(1-i\delta_n) \}$, $\Gamma$ is the gamma function, and $\{ \ket{g_{\mathrm{LZ}}^{(n)}(t)},\ket{e_{\mathrm{LZ}}^{(n)}(t)} \}$ is the instantaneous eigenbasis of $\hat{H}_{\mathrm{LZ}}^{(n)}(t)$. The evolution operator describing the Landau-Zener transition in the original representation is obtained with a simple back-transformation $\hat{N}^{(n)}(t_i-\Delta t,t_i+\Delta t) = \hat{U}_{\mathrm{LZ}}^{(n)}(t_i)\hat{N}_{\mathrm{LZ}}^{(n)}(t_i-\Delta t,t_i+\Delta t)(\hat{U}_{\mathrm{LZ}}^{(n)}(t_i))^{\dagger}$.

Since the system is in the charging regime, $\delta_1 \gg 1$ implying that $P_{\mathrm{LZ}}^{(1)} \ll 1$. Furthermore, $P_{\mathrm{LZ}}^{(n)}$ decreases with increasing $n$ since $\delta_n \sim \alpha_{n-1}^2$. Thus, with sufficiently large $n$, we are in the slow-passage limit \cite{pr492/1} and, additionally, the temporal suppression of the effective Hamiltonians in the adiabatic renormalization yields that $\ket{g_{\mathrm{LZ}}^{(n)}(t)} \approx \ket{0}$ and $\ket{e_{\mathrm{LZ}}^{(n)}(t)} \approx \ket{1}$ so that $\hat{N}^{(n)}(t_i-\Delta t,t_i+\Delta t)$ becomes approximately diagonal in the fixed basis. This implies that the transition causes an impulsive phase difference between the instantaneous eigenstates of the original effective Hamiltonian as
\begin{equation}
\begin{split}
&\Delta \Theta_{\mathrm{LZ}}^{(n)}(t_i-\Delta t,t_i+\Delta t) \\ &= \Theta_{\mathrm{LZ},g}^{(n)}(t_i-\Delta t,t_i+\Delta t) - \Theta_{\mathrm{LZ},e}^{(n)}(t_i-\Delta t,t_i+\Delta t) \\ &\approx 2\widetilde{\varphi}_S \\ &\approx -\pi.
\end{split}
\label{eq:res4}
\end{equation}
Note that in this limit, the transformation $\hat{U}_{\mathrm{LZ}}^{(n)}(t_i)$ becomes negligible  and, hence, the phase factor $\gamma_n(t_i)$ does not appear in the impulsive phase difference.

The reference times relevant to the interference are close to the transitions at $t_i-\Delta t$ and $t_{i+1} - \Delta t$, and the time-evolution of the system is described by the evolution operator $\hat{N}^{(n)}(t_i-\Delta t,t_i+\Delta t)\hat{U}^{(n)}(t_i + \Delta t, t_{i+1} - \Delta t)$, where $\hat{U}^{(n)}(t_{\mathrm{in}},t_{\mathrm{fin}})$ describes the adiabatic evolution in the original representation. Hence, the phase difference between the excitations is the sum of the impulsive and adiabatic phase differences accumulated between the states $\hat{D}_n(t)\ket{0}$ and $\hat{D}_n(t)\ket{1}$. This implies that the condition for the maximum constructive interference between the successive excitations resulting in the maximum transition probability is achieved with a frequency corresponding to
\begin{equation}
\begin{split}
&\Delta \Theta_{\mathrm{LZ}}^{(n)}(t_i-\Delta t,t_i + \Delta t) \\ &+ \Delta \Theta_T^{(n)}(t_i+\Delta t,t_{i+1}-\Delta t) \\ &= 2\pi N,
\end{split}
\label{eq:res5}
\end{equation}
where $N \in \mathbb{Z}$. Note that $\Delta \Theta_T^{(n)}(t_i+\Delta t,t_{i+1}-\Delta t) = \Delta \Theta_T^{(n)}(t_i,t_{i+1})$.

\begin{figure}
\begin{center}
\includegraphics[width=9cm]{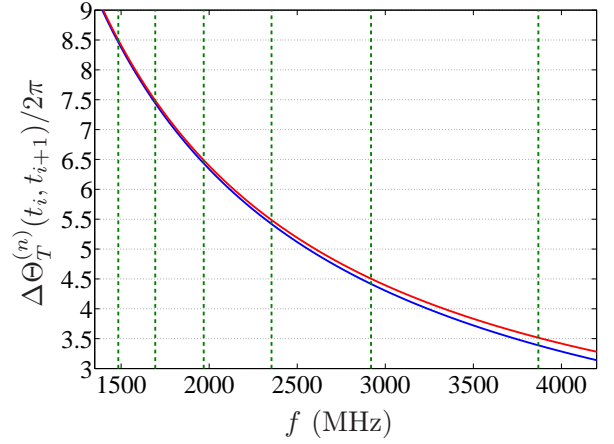}
\end{center}
\caption{\label{fig:exp1_res}(Color online) Comparison between the accumulated phase difference and the observed resonance peaks in the pumped current. The solid lines depict $\Delta \Theta_T^{(n)}(t_i,t_{i+1})$ for $n=1$ and $n=3$ from bottom to top. The dashed vertical lines are the resonance peak positions determined from the simulated pumped current in Fig.~\ref{fig:expIG}(b) for $g=0.025$.}
\end{figure}

In the limit of sufficiently large $n$, Eqs.~(\ref{eq:res4}) and (\ref{eq:res5}) yield an approximate condition for the frequencies corresponding to the downwards resonance peaks as
\begin{equation}
\begin{split}
\Delta \Theta_T^{(n)}(t_i,t_{i+1}) \approx 2\pi \left( N+\frac{1}{2} \right).
\end{split}
\label{eq:res6}
\end{equation}
Note that $\Delta \Theta_T^{(n)}(t_i,t_{i+1})$ is dependent on both the driving frequency and the used basis. The oscillations should only be present with high frequency and low environmental coupling strength as decoherence destroys any low-amplitude interference effects. We present the accumulated phase difference $\Delta \Theta_T^{(n)}(t_i,t_{i+1})$ and compare it with the observed peak positions in Fig.~\ref{fig:exp1_res}.
The resonance frequencies determined from the pumped current show an excellent agreement with the condition given in Eq.~(\ref{eq:res6}). This reaffirms our assumption on the origin of the oscillations as being caused by quantum interference between excitations and offers a way to predict the resonance behaviour in possible experiments. Using the adiabatic basis also allows for a good estimate for low frequencies. Furthermore, we conducted simulations similar to the ones in Fig.~\ref{fig:expIG} using $\varphi=0$ and obtained new downward resonance peak positions. Recalculating the accumulated phase difference allowed us to verify that the approximate condition in Eq.~(\ref{eq:res6}) still applies with excellent accuracy. Further results regarding quantum interference in the Cooper pair sluice, especially with respect to applications in phase interferometry, have been derived recently \cite{gasparinetti}.

Figures~\ref{fig:expIG}(a) and \ref{fig:expIG}(b), show that including dephasing decreases the amplitude of the oscillations. From the point-of-view of interference effects, this was to be expected as finite dephasing time implies the excitations to lose some of their phase coherence during the evolution. The current variation in the experimental data is significantly lower implying that the effective dephasing rate is most likely higher in the experimental setup than in our simulation. However, to account for the large variation in the gate voltage, a model accounting for a higher number of energy levels would need to be introduced resulting in more complex transitions. Thus, we refrain from making further comparisons and study the breaking point of the adiabaticity.

In an effort to suppress the oscillations, we perform the simulation for three different intervals for the gate charge maintaining $n_{\mathrm{range}} = 0.6$ and take an average over the results. Moving the midpoint of the gate voltage in our control cycle slightly from the degeneracy point alters the temporal dependence of the energy gap, and, thus changes the positions of the resonance peaks. Taking an average should then decrease the oscillations. We select the ranges as $n_g \in [0.2+\delta n_g;0.8+\delta n_g]$, where $\delta n_g \in \{-0.1,0,0.1\}$, so that the largest possible range for the gate variation is obtained while maintaining the two-state approximation with a reasonable accuracy. The result of such averaging procedure for $g=0.025$ with dephasing is shown in Fig.~\ref{fig:expIG}(c). In Fig.~\ref{fig:expIG}(c), we also give the pumped current with the same coupling strength and increased effective dephasing temperature to suppress the interference effects. Comparison with the experimental data shows that we obtain a decent estimate for the breakdown characteristics with this environmental coupling strength using both methods. The temporal dependence of the corresponding relaxation rate of the system $\Gamma_{eg} = |m_2^{(n)}|^2S(\omega_{01}^{(n)})$ and of the relaxation rate averaged over the three different simulations is presented in Fig.~\ref{fig:expIG}(d) at $f=14$ MHz. It is evident that the relaxation is strongest during gate operations due to non-adiabatic transitions, as expected.

\section{\label{sec:conclusions}Conclusions}

We introduced and demonstrated a method of applying successive coordinate transformations to describe accurately the dissipative dynamics of a steered two-level quantum system. Our method utilizes superadiabatic bases and the theory for nonsteered systems to obtain a master equation where the error resulting from the truncation of the perturbative expansion in the local adiabatic parameter is decreased.

We applied our method to Cooper pair pumping and showed that in the adiabatic limit, all orders of approximation return the ideal pumping result. In the zero-temperature limit and increased pumping frequency, increasing the strength of the environmental coupling was shown to induce ideal ground-state pumping only in the lowest order of our description. Furthermore, using high-order bases was shown to reduce the overestimation of the pumped charge stemming from the non-positivity of the reduced density matrix of the system. This is due to the high-order bases tracking the exact evolving state more closely. The major effect of these corrections was shown to be captured by the basis obtained with two transformations, i.e., the first superadiabatic basis.

We studied the breakdown of adiabaticity by simulating the pumped current with increasing pumping frequency. The high-order theory was shown to provide a more accurate picture of the robustness of the breakdown frequency against changes in the environment. An optimal strength of the environmental coupling was discovered preserving the adiabaticity of the system for the highest pumping frequency. The recently proposed \cite{prb82/134517} environment engineering scheme can potentially be used to probe and exploit this optimal point.

Finally, we applied our theory to model experimental pumping results similar to those of Ref.~\onlinecite{prl100/177201}. We altered the pumping frequency to simulate the increased pumping amplitude and observed oscillatory behaviour of the pumped current caused by quantum interference between driving-induced excitations generated at different times. We presented a condition for the highest excitation probability due to constructive interference and showed that the observed downward resonance peaks in the pumped current accurately corresponded to this condition. Using an averaging procedure and increasing the effective dephasing rate were methods that enabled us to finally present an estimate for the relaxation rate of the system. However, a many-state theory should be developed to facilitate more accurate predictions allowing us to alter the pumping amplitude. A seemingly valid approach could exploit the cyclic nature of the steering utilizing the Floquet theory, which would possibly allow for generalizations beyond the adiabatic evolution \cite{prb83/214508, arXiv:1108.3216}.

\begin{acknowledgments}

We thank J. J. Vartiainen and J. P. Pekola for co-providing the experimental findings. We also thank P. Solinas and S. Gasparinetti for helpful discussions. We acknowledge the Academy of Finland, the V\"ais\"al\"a Foundation, the KAUTE Foundation and the Emil Aaltonen Foundation for financial support. We have received funding from the European Community's Seventh Framework Programme under Grant Agreement No. 238345 (GEOMDISS).

\end{acknowledgments}

\bibliography{localbib.bib}

\begin{thebibliography}{10}%
\makeatletter
\providecommand \@ifxundefined [1]{%
 \ifx #1\undefined \expandafter \@firstoftwo
 \else \expandafter \@secondoftwo
\fi
}%
\providecommand \@ifnum [1]{%
 \ifnum #1\expandafter \@firstoftwo
 \else \expandafter \@secondoftwo
\fi
}%
\providecommand \enquote [1]{``#1''}%
\providecommand \bibnamefont  [1]{#1}%
\providecommand \bibfnamefont [1]{#1}%
\providecommand \citenamefont [1]{#1}%
\providecommand\href[0]{\@sanitize\@href}%
\providecommand\@href[1]{\endgroup\@@startlink{#1}\endgroup\@@href}%
\providecommand\@@href[1]{#1\@@endlink}%
\providecommand \@sanitize [0]{\begingroup\catcode`\&12\catcode`\#12\relax}%
\@ifxundefined \pdfoutput {\@firstoftwo}{%
 \@ifnum{\z@=\pdfoutput}{\@firstoftwo}{\@secondoftwo}%
}{%
 \providecommand\@@startlink[1]{\leavevmode\special{html:<a href="#1">}}%
 \providecommand\@@endlink[0]{\special{html:</a>}}%
}{%
 \providecommand\@@startlink[1]{%
  \leavevmode
  \pdfstartlink
   attr{/Border[0 0 1 ]/H/I/C[0 1 1]}%
   user{/Subtype/Link/A<</Type/Action/S/URI/URI(#1)>>}%
  \relax
 }%
 \providecommand\@@endlink[0]{\pdfendlink}%
}%
\providecommand \url  [0]{\begingroup\@sanitize \@url }%
\providecommand \@url [1]{\endgroup\@href {#1}{\urlprefix}}%
\providecommand \urlprefix [0]{URL }%
\providecommand \Eprint[0]{\href }%
\@ifxundefined \urlstyle {%
  \providecommand \doi [1]{doi:\discretionary{}{}{}#1}%
}{%
  \providecommand \doi [0]{doi:\discretionary{}{}{}\begingroup
  \urlstyle{rm}\Url }%
}%
\providecommand \doibase [0]{http://dx.doi.org/}%
\providecommand \Doi[1]{\href{\doibase#1}}%
\providecommand \bibAnnote [3]{%
  \BibitemShut{#1}%
  \begin{quotation}\noindent
    \textsc{Key:}\ #2\\\textsc{Annotation:}\ #3%
  \end{quotation}%
}%
\providecommand \bibAnnoteFile [2]{%
  \IfFileExists{#2}{\bibAnnote {#1} {#2} {\input{#2}}}{}%
}%
\providecommand \typeout [0]{\immediate \write \m@ne }%
\providecommand \selectlanguage [0]{\@gobble}%
\providecommand \bibinfo [0]{\@secondoftwo}%
\providecommand \bibfield [0]{\@secondoftwo}%
\providecommand \translation [1]{[#1]}%
\providecommand \BibitemOpen[0]{}%
\providecommand \bibitemStop [0]{}%
\providecommand \bibitemNoStop [0]{.\EOS\space}%
\providecommand \EOS [0]{\spacefactor3000\relax}%
\providecommand \BibitemShut [1]{\csname bibitem#1\endcsname}%
\bibitem{prsla392/45}%
  \BibitemOpen
  \bibfield{author}{%
  \bibinfo {author} {\bibfnamefont{M.~V.}\ \bibnamefont{Berry}},\ }%
  \bibfield{journal}{%
  \bibinfo {journal} {Proc. R. Soc. Lond. A}\ }%
  \textbf{\bibinfo {volume} {392}},\ \bibinfo {pages} {45} (\bibinfo {year}
  {1984})%
  \bibAnnoteFile{NoStop}{prsla392/45}%
\bibitem{prl51/2167}%
  \BibitemOpen
  \bibfield{author}{%
  \bibinfo {author} {\bibfnamefont{B.}~\bibnamefont{Simon}},\ }%
  \bibfield{journal}{%
  \bibinfo {journal} {Phys. Rev. Lett.}\ }%
  \textbf{\bibinfo {volume} {51}},\ \bibinfo {pages} {2167} (\bibinfo {year}
  {1983})%
  \bibAnnoteFile{NoStop}{prl51/2167}%
\bibitem{prl52/2111}%
  \BibitemOpen
  \bibfield{author}{%
  \bibinfo {author} {\bibfnamefont{F.}~\bibnamefont{Wilczek}}\ and\ \bibinfo
  {author} {\bibfnamefont{A.}~\bibnamefont{Zee}},\ }%
  \bibfield{journal}{%
  \bibinfo {journal} {Phys. Rev. Lett.}\ }%
  \textbf{\bibinfo {volume} {52}},\ \bibinfo {pages} {2111} (\bibinfo {year}
  {1984})%
  \bibAnnoteFile{NoStop}{prl52/2111}%
\bibitem{prl58/1593}%
  \BibitemOpen
  \bibfield{author}{%
  \bibinfo {author} {\bibfnamefont{Y.}~\bibnamefont{Aharonov}}\ and\ \bibinfo
  {author} {\bibfnamefont{J.}~\bibnamefont{Anandan}},\ }%
  \bibfield{journal}{%
  \bibinfo {journal} {Phys. Rev. Lett.}\ }%
  \textbf{\bibinfo {volume} {58}},\ \bibinfo {pages} {1593} (\bibinfo {year}
  {1987})%
  \bibAnnoteFile{NoStop}{prl58/1593}%
\bibitem{prl60/2339}%
  \BibitemOpen
  \bibfield{author}{%
  \bibinfo {author} {\bibfnamefont{J.}~\bibnamefont{Samuel}}\ and\ \bibinfo
  {author} {\bibfnamefont{R.}~\bibnamefont{Bhandari}},\ }%
  \bibfield{journal}{%
  \bibinfo {journal} {Phys. Rev. Lett.}\ }%
  \textbf{\bibinfo {volume} {60}},\ \bibinfo {pages} {2339} (\bibinfo {year}
  {1988})%
  \bibAnnoteFile{NoStop}{prl60/2339}%
\bibitem{pla264/94}%
  \BibitemOpen
  \bibfield{author}{%
  \bibinfo {author} {\bibfnamefont{P.}~\bibnamefont{Zanardi}}\ and\ \bibinfo
  {author} {\bibfnamefont{M.}~\bibnamefont{Rasetti}},\ }%
  \bibfield{journal}{%
  \bibinfo {journal} {Phys. Lett. A}\ }%
  \textbf{\bibinfo {volume} {264}},\ \bibinfo {pages} {94} (\bibinfo {year}
  {1999})%
  \bibAnnoteFile{NoStop}{pla264/94}%
\bibitem{nature407/355}%
  \BibitemOpen
  \bibfield{author}{%
  \bibinfo {author} {\bibfnamefont{G.}~\bibnamefont{Falci}}, \bibinfo {author}
  {\bibfnamefont{R.}~\bibnamefont{Fazio}}, \bibinfo {author}
  {\bibfnamefont{G.~M.}\ \bibnamefont{Palma}}, \bibinfo {author}
  {\bibfnamefont{J.}~\bibnamefont{Siewert}},\ and\ \bibinfo {author}
  {\bibfnamefont{V.}~\bibnamefont{Vedral}},\ }%
  \bibfield{journal}{%
  \bibinfo {journal} {Nature}\ }%
  \textbf{\bibinfo {volume} {407}},\ \bibinfo {pages} {355} (\bibinfo {year}
  {2000})%
  \bibAnnoteFile{NoStop}{nature407/355}%
\bibitem{science318/1889}%
  \BibitemOpen
  \bibfield{author}{%
  \bibinfo {author} {\bibfnamefont{P.~J.}\ \bibnamefont{Leek}}, \bibinfo
  {author} {\bibfnamefont{J.~M.}\ \bibnamefont{Fink}}, \bibinfo {author}
  {\bibfnamefont{A.}~\bibnamefont{Blais}}, \bibinfo {author}
  {\bibfnamefont{R.}~\bibnamefont{Bianchetti}}, \bibinfo {author}
  {\bibfnamefont{M.}~\bibnamefont{G\"oppl}}, \bibinfo {author}
  {\bibfnamefont{J.~M.}\ \bibnamefont{Gambetta}}, \bibinfo {author}
  {\bibfnamefont{D.~I.}\ \bibnamefont{Schuster}}, \bibinfo {author}
  {\bibfnamefont{L.}~\bibnamefont{Frunzio}}, \bibinfo {author}
  {\bibfnamefont{R.~J.}\ \bibnamefont{Schoelkopf}},\ and\ \bibinfo {author}
  {\bibfnamefont{A.}~\bibnamefont{Wallraff}},\ }%
  \bibfield{journal}{%
  \bibinfo {journal} {Science}\ }%
  \textbf{\bibinfo {volume} {318}},\ \bibinfo {pages} {1889} (\bibinfo {year}
  {2007})%
  \bibAnnoteFile{NoStop}{science318/1889}%
\bibitem{prb60/R9931}%
  \BibitemOpen
  \bibfield{author}{%
  \bibinfo {author} {\bibfnamefont{J.~P.}\ \bibnamefont{Pekola}}, \bibinfo
  {author} {\bibfnamefont{J.~J.}\ \bibnamefont{Toppari}}, \bibinfo {author}
  {\bibfnamefont{M.}~\bibnamefont{Aunola}}, \bibinfo {author}
  {\bibfnamefont{M.~T.}\ \bibnamefont{Savolainen}},\ and\ \bibinfo {author}
  {\bibfnamefont{D.~V.}\ \bibnamefont{Averin}},\ }%
  \bibfield{journal}{%
  \bibinfo {journal} {Phys. Rev. B}\ }%
  \textbf{\bibinfo {volume} {60}},\ \bibinfo {pages} {R9931} (\bibinfo {year}
  {1999})%
  \bibAnnoteFile{NoStop}{prb60/R9931}%
\bibitem{prb63/132508}%
  \BibitemOpen
  \bibfield{author}{%
  \bibinfo {author} {\bibfnamefont{M.}~\bibnamefont{Aunola}},\ }%
  \bibfield{journal}{%
  \bibinfo {journal} {Phys. Rev. B}\ }%
  \textbf{\bibinfo {volume} {63}},\ \bibinfo {pages} {132508} (\bibinfo {year}
  {2001})%
  \bibAnnoteFile{NoStop}{prb63/132508}%
\bibitem{prb68/020502(R)}%
  \BibitemOpen
  \bibfield{author}{%
  \bibinfo {author} {\bibfnamefont{M.}~\bibnamefont{Aunola}}\ and\ \bibinfo
  {author} {\bibfnamefont{J.~J.}\ \bibnamefont{Toppari}},\ }%
  \bibfield{journal}{%
  \bibinfo {journal} {Phys. Rev. B}\ }%
  \textbf{\bibinfo {volume} {68}},\ \bibinfo {pages} {020502(R)} (\bibinfo
  {year} {2003})%
  \bibAnnoteFile{NoStop}{prb68/020502(R)}%
\bibitem{prl91/177003}%
  \BibitemOpen
  \bibfield{author}{%
  \bibinfo {author} {\bibfnamefont{A.~O.}\ \bibnamefont{Niskanen}}, \bibinfo
  {author} {\bibfnamefont{J.~P.}\ \bibnamefont{Pekola}},\ and\ \bibinfo
  {author} {\bibfnamefont{H.}~\bibnamefont{Sepp\"a}},\ }%
  \bibfield{journal}{%
  \bibinfo {journal} {Phys. Rev. Lett.}\ }%
  \textbf{\bibinfo {volume} {91}},\ \bibinfo {pages} {177003} (\bibinfo {year}
  {2003})%
  \bibAnnoteFile{NoStop}{prl91/177003}%
\bibitem{prl95/256801}%
  \BibitemOpen
  \bibfield{author}{%
  \bibinfo {author} {\bibfnamefont{M.}~\bibnamefont{Governale}}, \bibinfo
  {author} {\bibfnamefont{F.}~\bibnamefont{Taddei}}, \bibinfo {author}
  {\bibfnamefont{R.}~\bibnamefont{Fazio}},\ and\ \bibinfo {author}
  {\bibfnamefont{F.~W.~J.}\ \bibnamefont{Hekking}},\ }%
  \bibfield{journal}{%
  \bibinfo {journal} {Phys. Rev. Lett.}\ }%
  \textbf{\bibinfo {volume} {95}},\ \bibinfo {pages} {256801} (\bibinfo {year}
  {2005})%
  \bibAnnoteFile{NoStop}{prl95/256801}%
\bibitem{apl90/082102}%
  \BibitemOpen
  \bibfield{author}{%
  \bibinfo {author} {\bibfnamefont{J.~J.}\ \bibnamefont{Vartiainen}}, \bibinfo
  {author} {\bibfnamefont{M.}~\bibnamefont{M\"ott\"onen}}, \bibinfo {author}
  {\bibfnamefont{J.~P.}\ \bibnamefont{Pekola}},\ and\ \bibinfo {author}
  {\bibfnamefont{A.}~\bibnamefont{Kemppinen}},\ }%
  \bibfield{journal}{%
  \bibinfo {journal} {App. Phys. Lett.}\ }%
  \textbf{\bibinfo {volume} {90}},\ \bibinfo {pages} {082102} (\bibinfo {year}
  {2007})%
  \bibAnnoteFile{NoStop}{apl90/082102}%
\bibitem{prl98/127001}%
  \BibitemOpen
  \bibfield{author}{%
  \bibinfo {author} {\bibfnamefont{M.}~\bibnamefont{Cholascinski}}\ and\
  \bibinfo {author} {\bibfnamefont{R.~W.}\ \bibnamefont{Chhajlany}},\ }%
  \bibfield{journal}{%
  \bibinfo {journal} {Phys. Rev. Lett.}\ }%
  \textbf{\bibinfo {volume} {98}},\ \bibinfo {pages} {127001} (\bibinfo {year}
  {2007})%
  \bibAnnoteFile{NoStop}{prl98/127001}%
\bibitem{prl100/027002}%
  \BibitemOpen
  \bibfield{author}{%
  \bibinfo {author} {\bibfnamefont{V.}~\bibnamefont{Brosco}}, \bibinfo {author}
  {\bibfnamefont{R.}~\bibnamefont{Fazio}}, \bibinfo {author}
  {\bibfnamefont{F.~W.~J.}\ \bibnamefont{Hekking}},\ and\ \bibinfo {author}
  {\bibfnamefont{A.}~\bibnamefont{Joye}},\ }%
  \bibfield{journal}{%
  \bibinfo {journal} {Phys. Rev. Lett.}\ }%
  \textbf{\bibinfo {volume} {100}},\ \bibinfo {pages} {027002} (\bibinfo {year}
  {2008})%
  \bibAnnoteFile{NoStop}{prl100/027002}%
\bibitem{prl100/117001}%
  \BibitemOpen
  \bibfield{author}{%
  \bibinfo {author} {\bibfnamefont{R.}~\bibnamefont{Leone}}, \bibinfo {author}
  {\bibfnamefont{L.~P.}\ \bibnamefont{L\'evy}},\ and\ \bibinfo {author}
  {\bibfnamefont{P.}~\bibnamefont{Lafarge}},\ }%
  \bibfield{journal}{%
  \bibinfo {journal} {Phys. Rev. Lett.}\ }%
  \textbf{\bibinfo {volume} {100}},\ \bibinfo {pages} {117001} (\bibinfo {year}
  {2008})%
  \bibAnnoteFile{NoStop}{prl100/117001}%
\bibitem{prb77/144522}%
  \BibitemOpen
  \bibfield{author}{%
  \bibinfo {author} {\bibfnamefont{S.}~\bibnamefont{Safaei}}, \bibinfo {author}
  {\bibfnamefont{S.}~\bibnamefont{Montangero}}, \bibinfo {author}
  {\bibfnamefont{F.}~\bibnamefont{Taddei}},\ and\ \bibinfo {author}
  {\bibfnamefont{R.}~\bibnamefont{Fazio}},\ }%
  \bibfield{journal}{%
  \bibinfo {journal} {Phys. Rev. B}\ }%
  \textbf{\bibinfo {volume} {77}},\ \bibinfo {pages} {144522} (\bibinfo {year}
  {2008})%
  \bibAnnoteFile{NoStop}{prb77/144522}%
\bibitem{prb73/214523}%
  \BibitemOpen
  \bibfield{author}{%
  \bibinfo {author} {\bibfnamefont{M.}~\bibnamefont{M\"ott\"onen}}, \bibinfo
  {author} {\bibfnamefont{J.~P.}\ \bibnamefont{Pekola}}, \bibinfo {author}
  {\bibfnamefont{J.~J.}\ \bibnamefont{Vartiainen}}, \bibinfo {author}
  {\bibfnamefont{V.}~\bibnamefont{Brosco}},\ and\ \bibinfo {author}
  {\bibfnamefont{F.~W.~J.}\ \bibnamefont{Hekking}},\ }%
  \bibfield{journal}{%
  \bibinfo {journal} {Phys. Rev. B}\ }%
  \textbf{\bibinfo {volume} {73}},\ \bibinfo {pages} {214523} (\bibinfo {year}
  {2006})%
  \bibAnnoteFile{NoStop}{prb73/214523}%
\bibitem{prl100/177201}%
  \BibitemOpen
  \bibfield{author}{%
  \bibinfo {author} {\bibfnamefont{M.}~\bibnamefont{M\"ott\"onen}}, \bibinfo
  {author} {\bibfnamefont{J.~J.}\ \bibnamefont{Vartiainen}},\ and\ \bibinfo
  {author} {\bibfnamefont{J.~P.}\ \bibnamefont{Pekola}},\ }%
  \bibfield{journal}{%
  \bibinfo {journal} {Phys. Rev. Lett.}\ }%
  \textbf{\bibinfo {volume} {100}},\ \bibinfo {pages} {177201} (\bibinfo {year}
  {2008})%
  \bibAnnoteFile{NoStop}{prl100/177201}%
\bibitem{prb81/174506}%
  \BibitemOpen
  \bibfield{author}{%
  \bibinfo {author} {\bibfnamefont{J.-M.}\ \bibnamefont{Pirkkalainen}},
  \bibinfo {author} {\bibfnamefont{P.}~\bibnamefont{Solinas}}, \bibinfo
  {author} {\bibfnamefont{J.~P.}\ \bibnamefont{Pekola}},\ and\ \bibinfo
  {author} {\bibfnamefont{M.}~\bibnamefont{M\"ott\"onen}},\ }%
  \bibfield{journal}{%
  \bibinfo {journal} {Phys. Rev. B}\ }%
  \textbf{\bibinfo {volume} {81}},\ \bibinfo {pages} {174506} (\bibinfo {year}
  {2010})%
  \bibAnnoteFile{NoStop}{prb81/174506}%
\bibitem{pra82/052304}%
  \BibitemOpen
  \bibfield{author}{%
  \bibinfo {author} {\bibfnamefont{P.}~\bibnamefont{Solinas}}, \bibinfo
  {author} {\bibfnamefont{J.-M.}\ \bibnamefont{Pirkkalainen}},\ and\ \bibinfo
  {author} {\bibfnamefont{M.}~\bibnamefont{M\"ott\"onen}},\ }%
  \bibfield{journal}{%
  \bibinfo {journal} {Phys. Rev. A}\ }%
  \textbf{\bibinfo {volume} {82}},\ \bibinfo {pages} {052304} (\bibinfo {year}
  {2010})%
  \bibAnnoteFile{NoStop}{pra82/052304}%
\bibitem{tToOQS}%
  \BibitemOpen
  \bibfield{author}{%
  \bibinfo {author} {\bibfnamefont{H.-P.}\ \bibnamefont{Breuer}}\ and\ \bibinfo
  {author} {\bibfnamefont{F.}~\bibnamefont{Pettrucione}},\ }%
  \emph{\bibinfo {title} {The Theory of Open Quantum Systems}}\ (\bibinfo
  {publisher} {Oxford University Press},\ \bibinfo {address} {Oxford},\
  \bibinfo {year} {2002})%
  \bibAnnoteFile{NoStop}{tToOQS}%
\bibitem{API}%
  \BibitemOpen
  \bibfield{author}{%
  \bibinfo {author} {\bibfnamefont{C.}~\bibnamefont{Cohen-Tannoudji}}, \bibinfo
  {author} {\bibfnamefont{J.}~\bibnamefont{Dupont-Roc}},\ and\ \bibinfo
  {author} {\bibfnamefont{G.}~\bibnamefont{Grynberg}},\ }%
  \emph{\bibinfo {title} {Atom-Photon Interactions}}\ (\bibinfo {publisher}
  {Wiley},\ \bibinfo {address} {New York},\ \bibinfo {year} {1992})%
  \bibAnnoteFile{NoStop}{API}%
\bibitem{prl94/070407}%
  \BibitemOpen
  \bibfield{author}{%
  \bibinfo {author} {\bibfnamefont{R.~S.}\ \bibnamefont{Whitney}}, \bibinfo
  {author} {\bibfnamefont{Y.}~\bibnamefont{Makhlin}}, \bibinfo {author}
  {\bibfnamefont{A.}~\bibnamefont{Shnirman}},\ and\ \bibinfo {author}
  {\bibfnamefont{Y.}~\bibnamefont{Gefen}},\ }%
  \bibfield{journal}{%
  \bibinfo {journal} {Phys. Rev. Lett.}\ }%
  \textbf{\bibinfo {volume} {94}},\ \bibinfo {pages} {070407} (\bibinfo {year}
  {2005})%
  \bibAnnoteFile{NoStop}{prl94/070407}%
\bibitem{pra73/052304}%
  \BibitemOpen
  \bibfield{author}{%
  \bibinfo {author} {\bibfnamefont{D.}~\bibnamefont{Parodi}}, \bibinfo {author}
  {\bibfnamefont{M.}~\bibnamefont{Sassetti}}, \bibinfo {author}
  {\bibfnamefont{P.}~\bibnamefont{Solinas}}, \bibinfo {author}
  {\bibfnamefont{P.}~\bibnamefont{Zanardi}},\ and\ \bibinfo {author}
  {\bibfnamefont{N.}~\bibnamefont{Zangh\`i}},\ }%
  \bibfield{journal}{%
  \bibinfo {journal} {Phys. Rev. A}\ }%
  \textbf{\bibinfo {volume} {73}},\ \bibinfo {pages} {052304} (\bibinfo {year}
  {2006})%
  \bibAnnoteFile{NoStop}{pra73/052304}%
\bibitem{pra73/022327}%
  \BibitemOpen
  \bibfield{author}{%
  \bibinfo {author} {\bibfnamefont{G.}~\bibnamefont{Florio}}, \bibinfo {author}
  {\bibfnamefont{P.}~\bibnamefont{Facchi}}, \bibinfo {author}
  {\bibfnamefont{R.}~\bibnamefont{Fazio}}, \bibinfo {author}
  {\bibfnamefont{V.}~\bibnamefont{Giovannetti}},\ and\ \bibinfo {author}
  {\bibfnamefont{S.}~\bibnamefont{Pascazio}},\ }%
  \bibfield{journal}{%
  \bibinfo {journal} {Phys. Rev. A}\ }%
  \textbf{\bibinfo {volume} {73}},\ \bibinfo {pages} {022327} (\bibinfo {year}
  {2006})%
  \bibAnnoteFile{NoStop}{pra73/022327}%
\bibitem{prl90/160402}%
  \BibitemOpen
  \bibfield{author}{%
  \bibinfo {author} {\bibfnamefont{A.}~\bibnamefont{Carollo}}, \bibinfo
  {author} {\bibfnamefont{I.}~\bibnamefont{Fuentes-Guridi}}, \bibinfo {author}
  {\bibfnamefont{M.~F.}\ \bibnamefont{Santos}},\ and\ \bibinfo {author}
  {\bibfnamefont{V.}~\bibnamefont{Vedral}},\ }%
  \bibfield{journal}{%
  \bibinfo {journal} {Phys. Rev. Lett.}\ }%
  \textbf{\bibinfo {volume} {90}},\ \bibinfo {pages} {160402} (\bibinfo {year}
  {2003})%
  \bibAnnoteFile{NoStop}{prl90/160402}%
\bibitem{prb77/115322}%
  \BibitemOpen
  \bibfield{author}{%
  \bibinfo {author} {\bibfnamefont{E.~M.}\ \bibnamefont{Gauger}}, \bibinfo
  {author} {\bibfnamefont{S.~C.}\ \bibnamefont{Benjamin}}, \bibinfo {author}
  {\bibfnamefont{A.}~\bibnamefont{Nazir}},\ and\ \bibinfo {author}
  {\bibfnamefont{B.~W.}\ \bibnamefont{Lovett}},\ }%
  \bibfield{journal}{%
  \bibinfo {journal} {Phys. Rev. B}\ }%
  \textbf{\bibinfo {volume} {77}},\ \bibinfo {pages} {115322} (\bibinfo {year}
  {2008})%
  \bibAnnoteFile{NoStop}{prb77/115322}%
\bibitem{pra82/052107}%
  \BibitemOpen
  \bibfield{author}{%
  \bibinfo {author} {\bibfnamefont{P.}~\bibnamefont{Pawlus}}\ and\ \bibinfo
  {author} {\bibfnamefont{E.}~\bibnamefont{Sj\"oqvist}},\ }%
  \bibfield{journal}{%
  \bibinfo {journal} {Phys. Rev. A}\ }%
  \textbf{\bibinfo {volume} {82}},\ \bibinfo {pages} {052107} (\bibinfo {year}
  {2010})%
  \bibAnnoteFile{NoStop}{pra82/052107}%
\bibitem{pra83/012112}%
  \BibitemOpen
  \bibfield{author}{%
  \bibinfo {author} {\bibfnamefont{P.}~\bibnamefont{Haikka}}, \bibinfo {author}
  {\bibfnamefont{J.~D.}\ \bibnamefont{Cresser}},\ and\ \bibinfo {author}
  {\bibfnamefont{S.}~\bibnamefont{Maniscalco}},\ }%
  \bibfield{journal}{%
  \bibinfo {journal} {Phys. Rev. A}\ }%
  \textbf{\bibinfo {volume} {83}},\ \bibinfo {pages} {012112} (\bibinfo {year}
  {2011})%
  \bibAnnoteFile{NoStop}{pra83/012112}%
\bibitem{pra81/022117}%
  \BibitemOpen
  \bibfield{author}{%
  \bibinfo {author} {\bibfnamefont{J.}~\bibnamefont{Hausinger}}\ and\ \bibinfo
  {author} {\bibfnamefont{M.}~\bibnamefont{Grifoni}},\ }%
  \bibfield{journal}{%
  \bibinfo {journal} {Phys. Rev. A}\ }%
  \textbf{\bibinfo {volume} {81}},\ \bibinfo {pages} {022117} (\bibinfo {year}
  {2010})%
  \bibAnnoteFile{NoStop}{pra81/022117}%
\bibitem{pra83/032122}%
  \BibitemOpen
  \bibfield{author}{%
  \bibinfo {author} {\bibfnamefont{J.~C.}\ \bibnamefont{Escher}}\ and\ \bibinfo
  {author} {\bibfnamefont{J.}~\bibnamefont{Ankerhold}},\ }%
  \bibfield{journal}{%
  \bibinfo {journal} {Phys. Rev. A}\ }%
  \textbf{\bibinfo {volume} {83}},\ \bibinfo {pages} {032122} (\bibinfo {year}
  {2011})%
  \bibAnnoteFile{NoStop}{pra83/032122}%
\bibitem{prl105/030401}%
  \BibitemOpen
  \bibfield{author}{%
  \bibinfo {author} {\bibfnamefont{J.~P.}\ \bibnamefont{Pekola}}, \bibinfo
  {author} {\bibfnamefont{V.}~\bibnamefont{Brosco}}, \bibinfo {author}
  {\bibfnamefont{M.}~\bibnamefont{M\"ott\"onen}}, \bibinfo {author}
  {\bibfnamefont{P.}~\bibnamefont{Solinas}},\ and\ \bibinfo {author}
  {\bibfnamefont{A.}~\bibnamefont{Shnirman}},\ }%
  \bibfield{journal}{%
  \bibinfo {journal} {Phys. Rev. Lett.}\ }%
  \textbf{\bibinfo {volume} {105}},\ \bibinfo {pages} {030401} (\bibinfo {year}
  {2010})%
  \bibAnnoteFile{NoStop}{prl105/030401}%
\bibitem{prb82/134517}%
  \BibitemOpen
  \bibfield{author}{%
  \bibinfo {author} {\bibfnamefont{P.}~\bibnamefont{Solinas}}, \bibinfo
  {author} {\bibfnamefont{M.}~\bibnamefont{M\"ott\"onen}}, \bibinfo {author}
  {\bibfnamefont{J.}~\bibnamefont{Salmilehto}},\ and\ \bibinfo {author}
  {\bibfnamefont{J.~P.}\ \bibnamefont{Pekola}},\ }%
  \bibfield{journal}{%
  \bibinfo {journal} {Phys. Rev. B}\ }%
  \textbf{\bibinfo {volume} {82}},\ \bibinfo {pages} {134517} (\bibinfo {year}
  {2010})%
  \bibAnnoteFile{NoStop}{prb82/134517}%
\bibitem{pra82/062112}%
  \BibitemOpen
  \bibfield{author}{%
  \bibinfo {author} {\bibfnamefont{J.}~\bibnamefont{Salmilehto}}, \bibinfo
  {author} {\bibfnamefont{P.}~\bibnamefont{Solinas}}, \bibinfo {author}
  {\bibfnamefont{J.}~\bibnamefont{Ankerhold}},\ and\ \bibinfo {author}
  {\bibfnamefont{M.}~\bibnamefont{M\"ott\"onen}},\ }%
  \bibfield{journal}{%
  \bibinfo {journal} {Phys. Rev. A}\ }%
  \textbf{\bibinfo {volume} {82}},\ \bibinfo {pages} {062112} (\bibinfo {year}
  {2010})%
  \bibAnnoteFile{NoStop}{pra82/062112}%
\bibitem{prsla414/31}%
  \BibitemOpen
  \bibfield{author}{%
  \bibinfo {author} {\bibfnamefont{M.~V.}\ \bibnamefont{Berry}},\ }%
  \bibfield{journal}{%
  \bibinfo {journal} {Proc. R. Soc. Lond. A}\ }%
  \textbf{\bibinfo {volume} {414}},\ \bibinfo {pages} {31} (\bibinfo {year}
  {1987})%
  \bibAnnoteFile{NoStop}{prsla414/31}%
\bibitem{GPIPberry}%
  \BibitemOpen
  \bibfield{author}{%
  \bibinfo {author} {\bibfnamefont{M.~V.}\ \bibnamefont{Berry}},\ }%
  \emph{\bibinfo {title} {\textnormal{in} Geometric Phases in Physics}}\
  (\bibinfo {publisher} {World Scientific},\ \bibinfo {address} {Singapore},\
  \bibinfo {year} {1988})%
  \bibAnnoteFile{NoStop}{GPIPberry}%
\bibitem{pra78/052508}%
  \BibitemOpen
  \bibfield{author}{%
  \bibinfo {author} {\bibfnamefont{G.}~\bibnamefont{Rigolin}}, \bibinfo
  {author} {\bibfnamefont{G.}~\bibnamefont{Ortiz}},\ and\ \bibinfo {author}
  {\bibfnamefont{V.~H.}\ \bibnamefont{Ponce}},\ }%
  \bibfield{journal}{%
  \bibinfo {journal} {Phys. Rev. A}\ }%
  \textbf{\bibinfo {volume} {78}},\ \bibinfo {pages} {052508} (\bibinfo {year}
  {2008})%
  \bibAnnoteFile{NoStop}{pra78/052508}%
\bibitem{ItS}%
  \BibitemOpen
  \bibfield{author}{%
  \bibinfo {author} {\bibfnamefont{M.}~\bibnamefont{Tinkham}},\ }%
  \emph{\bibinfo {title} {Introduction to Superconductivity: Second Edition}}\
  (\bibinfo {publisher} {Dover Publications},\ \bibinfo {address} {New York},\
  \bibinfo {year} {2004})%
  \bibAnnoteFile{NoStop}{ItS}%
\bibitem{rmp82/1155}%
  \BibitemOpen
  \bibfield{author}{%
  \bibinfo {author} {\bibfnamefont{A.~A.}\ \bibnamefont{Clerk}}, \bibinfo
  {author} {\bibfnamefont{M.~H.}\ \bibnamefont{Devoret}}, \bibinfo {author}
  {\bibfnamefont{S.~M.}\ \bibnamefont{Girvin}}, \bibinfo {author}
  {\bibfnamefont{F.}~\bibnamefont{Marquardt}},\ and\ \bibinfo {author}
  {\bibfnamefont{R.~J.}\ \bibnamefont{Schoelkopf}},\ }%
  \bibfield{journal}{%
  \bibinfo {journal} {Rev. Mod. Phys.}\ }%
  \textbf{\bibinfo {volume} {82}},\ \bibinfo {pages} {1155} (\bibinfo {year}
  {2010})%
  \bibAnnoteFile{NoStop}{rmp82/1155}%
\bibitem{prb71/012513}%
  \BibitemOpen
  \bibfield{author}{%
  \bibinfo {author} {\bibfnamefont{A.~O.}\ \bibnamefont{Niskanen}}, \bibinfo
  {author} {\bibfnamefont{J.~M.}\ \bibnamefont{Kivioja}}, \bibinfo {author}
  {\bibfnamefont{H.}~\bibnamefont{Sepp\"a}},\ and\ \bibinfo {author}
  {\bibfnamefont{J.~P.}\ \bibnamefont{Pekola}},\ }%
  \bibfield{journal}{%
  \bibinfo {journal} {Phys. Rev. B}\ }%
  \textbf{\bibinfo {volume} {71}},\ \bibinfo {pages} {012513} (\bibinfo {year}
  {2005})%
  \bibAnnoteFile{NoStop}{prb71/012513}%
\bibitem{ajp66/431}%
  \BibitemOpen
  \bibfield{author}{%
  \bibinfo {author} {\bibfnamefont{G.~G.}\ \bibnamefont{de~Polavieja}}\ and\
  \bibinfo {author} {\bibfnamefont{E.}~\bibnamefont{Sj\"oqvist}},\ }%
  \bibfield{journal}{%
  \bibinfo {journal} {Am. J. Phys.}\ }%
  \textbf{\bibinfo {volume} {66}},\ \bibinfo {pages} {431} (\bibinfo {year}
  {1998})%
  \bibAnnoteFile{NoStop}{ajp66/431}%
\bibitem{pr492/1}%
  \BibitemOpen
  \bibfield{author}{%
  \bibinfo {author} {\bibfnamefont{S.~N.}\ \bibnamefont{Shevchenko}}, \bibinfo
  {author} {\bibfnamefont{S.}~\bibnamefont{Ashhab}},\ and\ \bibinfo {author}
  {\bibfnamefont{F.}~\bibnamefont{Nori}},\ }%
  \bibfield{journal}{%
  \bibinfo {journal} {Phys. Rep.}\ }%
  \textbf{\bibinfo {volume} {492}},\ \bibinfo {pages} {1} (\bibinfo {year}
  {2010})%
  \bibAnnoteFile{NoStop}{pr492/1}%
\bibitem{prsla137/696}%
  \BibitemOpen
  \bibfield{author}{%
  \bibinfo {author} {\bibfnamefont{C.}~\bibnamefont{Zener}},\ }%
  \bibfield{journal}{%
  \bibinfo {journal} {Proc. R. Soc. Lond. A}\ }%
  \textbf{\bibinfo {volume} {137}},\ \bibinfo {pages} {696} (\bibinfo {year}
  {1932})%
  \bibAnnoteFile{NoStop}{prsla137/696}%
\bibitem{gasparinetti}%
  \BibitemOpen
  \bibfield{author}{%
  \bibinfo {author} {\bibfnamefont{S.}~\bibnamefont{Gasparinetti}}, \bibinfo
  {author} {\bibfnamefont{P.}~\bibnamefont{Solinas}},\ and\ \bibinfo {author}
  {\bibfnamefont{J.~P.}\ \bibnamefont{Pekola}},\ }%
  \bibinfo {journal} {arXiv:1106.3941 (unpublished)}%
  \bibAnnoteFile{NoStop}{gasparinetti}%
\bibitem{prb83/214508}%
  \BibitemOpen
\bibfield{journal}{%
    }%
  \bibfield{author}{%
  \bibinfo {author} {\bibfnamefont{A.}~\bibnamefont{Russomanno}}, \bibinfo
  {author} {\bibfnamefont{S.}~\bibnamefont{Pugnetti}}, \bibinfo {author}
  {\bibfnamefont{V.}~\bibnamefont{Brosco}},\ and\ \bibinfo {author}
  {\bibfnamefont{R.}~\bibnamefont{Fazio}},\ }%
  \bibfield{journal}{%
  \bibinfo {journal} {Phys. Rev. B}\ }%
  \textbf{\bibinfo {volume} {83}},\ \bibinfo {pages} {214508} (\bibinfo {year}
  {2011})%
  \bibAnnoteFile{NoStop}{prb83/214508}%
\bibitem{arXiv:1108.3216}%
  \BibitemOpen
  \bibfield{author}{%
  \bibinfo {author} {\bibfnamefont{I.}~\bibnamefont{Kamleitner}}\ and\ \bibinfo
  {author} {\bibfnamefont{A.}~\bibnamefont{Shnirman}},\ }%
  \bibinfo {journal} {arXiv:1108.3216 (unpublished)}%
  \bibAnnoteFile{NoStop}{arXiv:1108.3216}%
\end{thebibliography}%

\end{document}